\documentclass[prd,preprint,eqsecnum,nofootinbib,amsmath,amssymb,
               tightenlines,dvips]{revtex4}
\usepackage{graphicx}
\usepackage{bm}

\begin {document}


\def\Mrowczynski{Mr\'owczy\'nski}
\def\Qs{Q_{\rm s}}

\def\half{{\textstyle{\frac12}}}
\def\p{{\bm p}}
\def\q{{\bm q}}
\def\x{{\bm x}}
\def\v{{\bm v}}
\def\E{{\bm E}}
\def\B{{\bm B}}
\def\A{{\bm A}}
\def\a{{\bm a}}
\def\b{{\bm b}}
\def\c{{\bm c}}
\def\j{{\bm j}}
\def\n{{\bm n}}
\def\grad{{\bm\nabla}}
\def\da{d_{\rm A}}
\def\tr{\operatorname{tr}}
\def\Im{\operatorname{Im}}
\def\md{m_{\rm D}}
\def\mpl{m_{\rm pl}}
\def\pol{\varepsilon}
\def\bpol{{\bm\pol}}



\title
    {
    QCD Plasma Instabilities and Bottom-Up Thermalization
    }

\author {Peter Arnold}
\author {Jonathan Lenaghan}
\affiliation
    {%
    Department of Physics,
    University of Virginia,
    P.O. Box 400714
    Charlottesville, VA 22901-4714
    }%
\author{Guy D. Moore}
\affiliation
    {%
    Department of Physics,
    McGill University, 
    3600 University St.,
    Montr\'{e}al, QC H3A 2T8, Canada
    }%

\date {February 2003}

\begin {abstract}%
    {%
    We study 
    the role of QCD plasma instabilities
    in non-equilibrium quark-gluon plasmas.  First, we argue that such
    instabilities must drastically modify the ``bottom-up'' thermalization
    scenario for heavy-ion collisions.  Second,
    we discuss conditions for the existence of instabilities in a more
    general context than previously treated in the QCD literature.  We
    also give a thorough qualitative review of the origin of instabilities.
    We discuss some mechanisms whereby the growth of plasma instabilities
    saturates.
    Finally, we solve explicitly for instabilities and their growth rates
    for two extreme cases of anisotropic non-equilibrium plasmas that can be
    treated relatively simply and analytically:
    $f(\p) = F(p_\perp) \, \delta(p_z)$ and
    $f(\p) = F(p_z) \, \delta^{(2)}(p_\perp)$,
    where $f(\p)$ is the distribution of particles in momentum
    space.
    }%
\end {abstract}

\maketitle
\thispagestyle {empty}


\section {Introduction}
\label{sec:intro}

\Mrowczynski\ \cite{mrow0,mrow1,mrow2,mrow3,mrow&thoma,randrup&mrow}
has long advocated that plasma instabilities may play
an important role in the thermalization of quark-gluon plasmas from
anisotropic, non-equilibrium initial conditions, such as those
pertaining to heavy-ion collisions.
(For other early discussions, see
Refs.\ \cite{heinz_conf,selikhov1,selikhov2,selikhov3,pavlenko}.)
Since heavy-ion collisions are rather complicated, with a large
variety of different physical processes playing roles at different
stages of the collision, it is useful to examine this
proposal in a theoretically clean limit.
A natural framework for such an investigation has been provided
by Baier, Mueller, Schiff, and Son \cite{BMSS}.
In a beautiful paper, they consider what happens shortly after the
collision in the saturation scenario \cite{gribov,blaizot,qiu,larry},
at energies high enough that the effective strong
coupling $\alpha$ at the relevant momentum scale $\Qs$
(known as the saturation scale) is weak,
$\alpha(\Qs) \ll 1$.
The result of their analysis, which found an interesting variety
of parametrically different time scales
characteristic of different stages of thermalization
(such as $\alpha^{-3/2} \Qs^{-1}$,
$\alpha^{-5/2} \Qs^{-1}$, and $\alpha^{-13/5} \Qs^{-1}$),
is known as bottom-up thermalization.
We will show that even the very first moments of this
scenario are drastically modified by the effects of plasma instabilities,
which were not considered by Baier {\it et al.}

We will use collisionless kinetic theory
coupled to soft classical gauge fields (also known as the Hard Thermal Loop
approximation) to study the plasma instabilities
during the first phase of bottom-up thermalization.
First, in section \ref{sec:BMSS}, 
we will briefly review the first
phase of the bottom-up thermalization scenario and summarize our
argument that it must be drastically modified.
We also argue that the collisionless approximation is appropriate
to check for instabilities.
For those readers more
interested in the plasma instabilities of kinetic theory than in the
details of the bottom-up thermalization scenario, the upshot is that
we will be motivated to study the existence and properties of
plasma instabilities for an initial, homogeneous, non-equilibrium,
phase-space distribution of hard gluons of approximately
the following form:
\begin {equation}
   f(\p,\x) = F(p_\perp) \, \delta(p_z) ,
\label {eq:fplanar}
\end {equation}
where $p_\perp \equiv \sqrt{p_x^2+p_y^2}$.
This is an interesting limiting case of certain
classes of distributions that have been
studied in the literature numerically
by Randrup and \Mrowczynski\ \cite{randrup&mrow} and
Romatschke and Strickland \cite{strickland}, but (\ref{eq:fplanar})
has the advantage of being more analytically tractable.%
\footnote{
   Romatschke and Strickland \cite{strickland}
   do obtain analytic results in a more
   general class of situations, but their results are unwieldy enough
   that they refrain from giving them explicitly in their paper.
   Also, another example of analysis of instabilities in ultra-relativistic
   plasmas, generalized to inclusion of a background magnetic field but
   restricted to wave vectors in certain symmetry directions, may be found in
   Ref.\ \cite{yang}.
}
The particular plasma instabilities that will be of interest are known
in various contexts as
(i) Weibel \cite{weibel}, filamentation, or pinching instabilities,
and (ii) two-stream or Buneman \cite{buneman} instabilities,
with the names somewhat interchangeable in each category.
Because many readers will not be intimately familiar with these
instabilities in kinetic theory, we will review them both
formally and qualitatively in section \ref{sec:2stream}
for parity-invariant situations.
Typically, for parity-invariant particle momentum distributions,
plasma instabilities exist (in the collisionless approximation)
whenever those distributions are anisotropic.  We will give
conditions which make this statement precise.
We also discuss certain aspects of the 
physics that eventually cuts off the growth of such 
instabilities and point out the qualitative differences 
between the cases of Abelian and 
non-Abelian gauge theories.  To make the rather general 
discussion in section \ref{sec:2stream} more concrete, 
we will then analyze in detail in section
\ref{sec:planar} the instabilities associated with the particle
distribution (\ref{eq:fplanar}).
This distribution is simple enough that much of the analysis can be done
analytically.
We also give a specific example in section \ref{sec:twirl} of how
non-Abelian interactions can cut off the growth of plasma instabilities.
Various matters are left for appendices, including an analytic analysis of
instabilities for the distribution
\begin {equation}
   f(\p,\x) = F(p_z) \, \delta^{(2)}(\p_\perp) ,
\label {eq:fline}
\end {equation}
which is complementary to that of (\ref{eq:fplanar}).

We will show that plasma instabilities play a crucial role in the
post-saturation evolution of heavy-ion collisions, in the framework
originally established for bottom-up thermalization,
but we are not yet able to give detailed parametric estimates of
the time scale for equilibration
(analogous to the $\alpha^{-13/5} \Qs^{-1}$ time scale
of Baier {\it et al.}).  We do expect thermalization to be more rapid,
but this is a topic for future research.


\subsection {The first phase of bottom-up thermalization}
\label{sec:BMSS}

\begin{figure}
\includegraphics[scale=0.40]{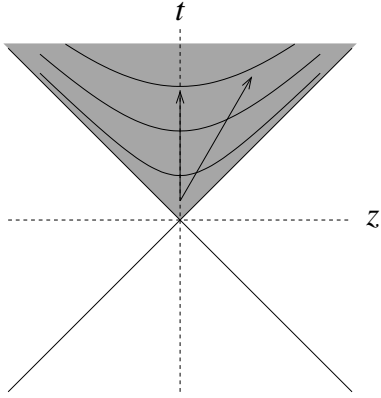}
\caption{%
    The standard picture of lines of constant proper time
    $\tau = \sqrt{t^2-z^2}$ in a high-energy collision, in which the
    $45^\circ$ lines are the colliding nuclei.
    The two arrows show how free-streaming partons leaving from
    the same initial point would later segregate themselves in $z$
    according to their value of $v_z$.
    \label{fig:tau}
    }
\end{figure}

Suppose the initial nuclei are moving in the $\pm z$ directions.
The starting point for the bottom-up thermalization scenario is to consider
an initial distribution of particles that is (i) boost-invariant in
the $z$ direction, (ii) homogeneous in the $x$ and $y$ directions,
and (iii) dominated by gluons
with momenta $p$ of order some scale $\Qs$
and with non-perturbative occupation numbers per mode
of order $1/\alpha$.
For sufficiently high energy collisions, the first
assumption should be adequate for describing the central rapidity region
and the second assumption for describing local thermalization
for large nuclei.
These two assumptions imply that physical
quantities (measured in a local rest frame)
depend only on the proper time $\tau = \sqrt{t^2-z^2}$,
as depicted in Fig.\ \ref{fig:tau}.
The third assumption is the saturation assumption and is taken to
hold at $\tau \sim \Qs^{-1}$.  The bottom-up thermalization scenario
concerns itself with what happens at later times $\tau \gg \Qs^{-1}$,
given this initial condition.  Following Baier {\it et al.},
we will refer to the initial gluons as ``hard'' gluons.%
\footnote
    {%
    This usage is to distinguish them from lower energy partons
    generated later in the evolution of the collision.  From
    the point of view of
    the colored glass condensate picture of the nucleus, or of 
    ``mini-jet'' analyses of heavy ion collisions, the usage may seem
    peculiar, since $\Qs$ is treated as an infrared cutoff in those
    contexts.%
    }

If there were no interactions, then, as time progressed,
the hard gluons would simply follow
straight-line trajectories and segregate themselves in $z$ according to
their values of $v_z = p_z/p$, as depicted in Fig.\ \ref{fig:tau}.
One can show that, in a local rest frame (where the average $\p$ vanishes),
the resulting distribution of hard gluons for $\tau \gg \Qs^{-1}$
would then be dominated by gluons with
$p_x \sim p_y \sim \Qs$ but $p_z \sim 1/\tau \ll \Qs$.
The occupation numbers of these modes would still be of order
$1/\alpha$.  The spatial number density $N_{\rm h}$ of hard gluons
decreases linearly with the linear expansion as
\begin {equation}
   N_{\rm h} \sim \frac{\Qs^3}{\alpha(\Qs\tau)} .
\label {eq:Nh}
\end {equation}

Baier {\it et al.}\ argue that small-angle scattering between
the hard gluons instead widens the spread in $p_z$ to
\begin {equation}
   p_z \sim \frac{\Qs}{(\Qs\tau)^{1/3}}
\label{eq:pz}
\end {equation}
during the first stage, $1 \ll \Qs\tau \ll \alpha^{-3/2}$, of their scenario.
They then argue that hard gluon number is conserved over the short duration
of this first stage, and so
the spatial number density of hard gluons still decreases linearly with
$\tau$ as (\ref{eq:Nh}).  The occupation number per mode
(related by $N_{\rm h} = \int_\p f_{\rm h} \sim p_x p_y p_z f_{\rm h}$)
then decreases as
\begin {equation}
   f_{\rm h} \sim \frac{N_{\rm h}}{\Qs^2 p_z}
   \sim \frac{1}{\alpha (\Qs\tau)^{2/3}} .
\end {equation}
The fact that $f_{\rm h} \ll 1/\alpha$ for $\Qs\tau \gg 1$ means that
the direct interactions between hard particles can be treated perturbatively
at these times.

Unscreened Coulomb scattering is infrared divergent.
The small-angle scattering of hard gluons discussed above is
therefore dominated by momentum exchanges
of order the inverse screening length.
Baier {\it et al.}\ estimated this soft
scale, which we will call  $m_{\rm soft}$,
to be%
\footnote{
  Our terminology is slightly different from that of Baier
  {\it et al.}\ \cite{BMSS}.  They refer to (\ref{eq:pz})
  as the ``soft'' physics
  scale during the first stage of their scenario.
  In our paper, the term ``soft'' will instead generally refer to
  (\ref{eq:msoft}).
}
\begin {equation}
   m_{\rm soft}^2 \sim \alpha \int_\p \frac{f_{\rm h}(\p)}{p}
   \sim \frac{\alpha N_{\rm h}}{\Qs} \sim \frac{\Qs}{\tau} .
\label{eq:msoft}
\end {equation}
They loosely referred to $m_{\rm soft}$ as the Debye mass $m_{\rm D}$
and took estimates from the literature \cite{debye1,debye2,debye3}.
In fact, the
Debye mass is not precisely the correct quantity to use since one really
wants to use the full self-energy $\Pi^{\mu\nu}(\omega,\q)$ for the
soft exchanged gluon of the small-angle scattering process.
Also, the Debye mass is not precisely defined in anisotropic situations:
it depends on direction.
However, these details do not affect parametric
estimates, and so (\ref{eq:msoft}) is a perfectly adequate
estimate of the relevant soft physics scale.  There is only one problem.
A parametric estimate does not necessarily reveal the {\em sign}\/
of the result.  As we shall see, a more appropriate definition of
$m_{\rm soft}^2$ reveals that its sign is negative for certain modes,
corresponding to an instability in the soft gauge fields in the presence
of the anisotropic distribution of hard gluons.  These unstable soft
modes then grow exponentially with a characteristic time scale of
\begin {equation}
   t_{\rm growth} \sim \frac1{|m_{\rm soft}|}
   \sim \sqrt{\frac{\tau}{\Qs}}.
\label {eq:tgrowth}
\end {equation}
Because the growth time $t_{\rm growth}$ is parametrically small compared
to the expansion time $\tau$ for $\Qs\tau \gg 1$,
the first stage of bottom-up thermalization is drastically modified
by the generation of large, non-perturbative soft gauge fields.
Since this is true for {\em any}\/ $\Qs\tau \gg 1$, the first stage
of the original bottom-up thermalization scenario cannot, even briefly,
be qualitatively accurate.

In order to demonstrate the instability, note that the first stage
of the bottom-up scenario is described by
hard gluons with $p_z \ll p_x \sim p_y \sim \Qs$ as in (\ref{eq:pz}).
Using the azimuthal symmetry of the problem, such a distribution of
hard gluons can be approximately described as having
the form (\ref{eq:fplanar}).  In Sec.\ \ref{sec:planar}, we will analyze
the soft instabilities created by hard distributions of this form.

The extreme planarity (oblateness) of the anisotropy of our
model distribution (\ref{eq:fplanar}) is somewhat an oversimplification of the
actual situation in the bottom-up scenario.  Baier {\it et al.}\ point out that
not only hard gluons contribute to $m_{\rm soft}^2$, but so also do
softer gluons which have been generated by Bremsstrahlung and then pushed
to momenta $k_{\rm s}$ of order $p_z$ (\ref{eq:pz}) by scattering from
hard particles.  These softer gluons will be anisotropic but not
have the extreme anisotropy of the hard gluons.  In the bottom-up
thermalization scenario, Baier {\it et al.}\ show that such soft gluons give a
contribution to $m_{\rm soft}^2$ comparable to that
of the hard gluons.  We will see that
whenever there is significant deviation from anisotropy, then there are
unstable modes with momenta of order $m_{\rm soft}$, and so plasma
instabilities will indeed modify the bottom-up scenario.  However,
as a simple analytically-tractable example of this phenomena, it will be
instructive to ignore the softer contribution and study the effects
of just the extreme oblate distribution (\ref{eq:fplanar}) of
the hard gluons.

In our analysis of plasma instabilities and their growth, we will use a
collisionless approximation for the hard particles.  We need to justify
this approximation since a significant rate of collisions could eliminate
the instability.  Baier {\it et al.}\ estimate the rate of hard particle
collisions for a single hard gluon in the first stage of their scenario as
\begin {equation}
   \frac{dN_{\rm col}}{d\tau} \sim \sigma N_{\rm h} (1+f_{\rm h})
   \sim \frac{\alpha N_{\rm h}}{m_{\rm soft}^2 p_z \tau} ,
\end {equation}
where $\sigma \sim \alpha^2 m_{\rm soft}^2$ is the cross-section.
Using the previous estimates for $N_{\rm h}$, $m_{\rm soft}$, and
$p_z$, the mean free time between such collisions is then
\begin {equation}
   t_{\rm col} \sim
   \left(\frac{dN_{\rm col}}{d\tau}\right)^{-1}
   \sim \frac{(\Qs\tau)^{2/3}}{\Qs} .
\end {equation}
For $\Qs\tau \gg 1$,
this is parametrically long compared to the time scale $t_{\rm growth}$
of (\ref{eq:tgrowth}) for the growth of instabilities.
We may therefore ignore collisions of
hard gluons for the purpose of understanding whether the bottom-up
thermalization scenario is affected by plasma instabilities.

In our analysis of instabilities, we will assume that the plasma
is homogeneous.  In fact, from Fig.\ \ref{fig:tau}, one can see that
densities will vary in a constant $t$ slice over a distance scale
of order $\tau$.  The homogeneity assumption is valid for
$\Qs\tau \gg 1$ because this scale will be large compared to the
typical wavelength $1/|m_{\rm soft}|$ of unstable modes.


\section {Plasma instabilities}
\label{sec:2stream}

The basic starting point is the expression for the soft gauge-field
self-energy $\Pi^{\mu\nu}(\omega,\q)$
in the presence of a given homogeneous but
non-equilibrium distribution $f(\p)$ of hard particles.  In this
section, we will treat general distributions $f(\p)$ and will not yet
specialize to specific forms such as the planar distribution of
(\ref{eq:fplanar}).


\subsection {Review of the retarded self-energy}

First, let us briefly review the known result for the self-energy
from collisionless kinetic theory.  In order to retain contact with
the literature on non-relativistic plasmas, we will
avoid specializing to the ultra-relativistic limit
for the moment.  For simplicity,
we start by discussing Abelian gauge theories such as
electromagnetism.
One starts
with the Vlasov equations, which are the collisionless Boltzmann equation
for a hard particle distribution $f(\p,\x,t)$
and Maxwell's equations for the soft gauge fields:
\begin {equation}
   \partial_t f + \v\cdot\grad_\x f + e (\E + \v\times\B) \cdot \grad_\p f
   = 0 ,
\end {equation}
\begin {equation}
   \partial_\nu F^{\mu\nu} = j^\mu = e \int_\p v^\mu f ,
\label {eq:Maxwell}
\end {equation}
where $e(\E+\v\times\B)$ is the Lorentz force on hard particles of
charge $e$, $\v = \v(\p)$ is the hard particle velocity for a given
momentum $\p$, and $v^\mu \equiv (1,\v)$.
There is also an implicit sum over particle species and their charges on
the right-hand side of (\ref{eq:Maxwell}).
In the absence of external fields, one 
treats $\E(\x,t)$ and $\B(\x,t)$ as small
and linearizes
$f(\p,\x,t)=f_0(\p)+f_1(\p,\x,t)$
in small fluctuations $f_1$ about some homogeneous distribution $f_0(\p)$.
We will assume that $e \int_\p v^\mu f_0$ vanishes (when implicitly
summed over species) so that $f_0$ carries no net charge or net current.
Then
\begin {equation}
   \partial_t f_1 + \v\cdot\grad_\x f_1
             + e (\E + \v\times\B) \cdot \grad_\p f_0
   \simeq 0 ,
\end {equation}
\begin {equation}
   \partial_\nu F^{\mu\nu} = e \int_\p v^\mu f_1 ,
\end {equation}
which are known as linearized Vlasov equations.
Fourier transforming from $(t,\x)$ to $(\omega,\q)$, one may solve
for $f_1$ and obtain
\begin {equation}
   iQ_\nu F^{\mu\nu}
   \simeq i e^2 \int_\p \frac{v^\mu (\E+\v\times\B)\cdot\grad_\p f_0 }
                           {-\omega+\v\cdot\q-i\epsilon}
        \, ,
\label {eq:Maxwell2}
\end {equation}
where the $\epsilon$ is a positive infinitesimal inserted as a
prescription to obtain the retarded solution.
We use the shorthand notation
\begin {equation}
   \int_\p \cdots \quad \equiv \quad \int \frac{d^3p}{(2\pi)^3} \, \cdots .
\end {equation}
We may rewrite (\ref{eq:Maxwell2}) as
\begin {equation}
   i Q_\nu F^{\mu\nu} \simeq - \Pi^{\mu\nu} A_\nu ,
\label {eq:Maxwell3}
\end {equation}
extracting the result for the retarded self-energy $\Pi^{\mu\nu}$,
\begin {equation}
   \Pi^{\mu\nu}(\omega,\q) =
   e^2 \int_\p
     \frac{\partial f(\p)}{\partial p^k}
     \left[ -v^\mu g^{k\nu}
            + \frac{v^\mu v^\nu q^k}{-\omega+\v\cdot\q-i\epsilon} \right] ,
\label {eq:pi}
\end {equation}
where we use $({-}{+}{+}{+})$ metric convention, and
Roman indices such as $k$ are spatial indices that run from 1 to 3.
The self-energy $\Pi^{\mu\nu}$ is symmetric in the indices $\mu\nu$.%
\footnote{
  The symmetry is not obvious for
  the first term in (\ref{eq:pi}) unless one integrates by parts
  and uses $\v = \grad_\p E_\p$ so that
  $\partial_{p^j} v^i = \partial_{p^j} \partial_{p^i} E_\p$,
  where $E_\p$ is the energy of a hard particle with momentum $\p$.
}
It is also useful to note that the above expression satisfies the
Ward identity, $Q_\mu \Pi^{\mu\nu} = 0$ (which is just
current conservation, $Q_\mu j^\mu = 0$),
where $Q \equiv (\omega,\q)$.
This implies
\begin {equation}
   \Pi^{0\nu} = \frac{q^i \Pi^{i\nu}}{\omega} \,,
   \qquad
   \Pi^{00} = \frac{q^i \Pi^{ij} q^j}{\omega^2} .
\label{eq:ward}
\end {equation}

The result for non-Abelian gauge theory%
\footnote{
  See Ref.\ \cite{mrow&thoma} for the non-Abelian version of
  (\ref{eq:pi}) in the non-equilibrium case.  However, the
  basic modifications have been known in various approximations for
  ages ({\it e.g.}\ Ref.\ \cite{heinz}).  Ref.\ \cite{birse} also 
  has expressions for (\ref{eq:pi}) specialized to the case of 
  axial symmetry.  
}
is essentially identical
except that $e^2$ is replaced by
$g^2 \tr(T_s^a T_s^b) = \delta^{ab} C_s d_s/\da$
for each species $s$ of hard particle, where $g$ is the gauge coupling,
$a$ and $b$ are adjoint color indices for the soft gluon,
$T_s^a$ are the color generators for the hard particle's color
representation, $C_s$ is the quadratic
Casimir given by $T_s^a T_s^a = C_s$, and $d_s$, $\da$ are the
dimensions of the particle's representation and the adjoint
representation respectively.  For gluons in QCD, $C_s d_s/\da = 3$,
while for quarks $C_s d_s/\da = 1/2$.
However, since there is no essential difference between the Abelian and
non-Abelian results of the linearized analysis, we will generally write
the self-energy as (\ref{eq:pi}) for the sake of brevity, with
the understanding that
\begin {equation}
   e^2 \to \sum_s \nu_s \frac{g^2 C_s d_s}{\da}
\end {equation}
with $f \to f_s$,
where 
$\nu_s$ represents the number of degrees of freedom of a given
hard particle species,
excluding color.
(So, for instance, if $f_{\rm g}$ for gluons represents the density of gluons
{\em per}\/ spin state and color, then $\nu_{\rm g}$ would be 2.)

Isotropic distributions $f(\p)=f(p)$ have the following special properties:
(i) $\Pi(0,\q)$ is diagonal with the only non-zero entry
$\Pi^{00}(0,\q) = -\md^2$, where $\md$ is the Debye screening mass
(inverse Debye screening length); and
(ii) $\Pi(\omega,0)$ is diagonal with the only non-zero entries
$\Pi^{ij}(\omega,0) = \mpl^2 \delta^{ij}$, where
$\mpl = \md/\sqrt3$ is the plasmon mass (plasma frequency).
In contrast, for generic anisotropic distributions,
(i) $\Pi(0,\q)$ is not diagonal and depends on the direction $\hat\q$,
and (ii) $\Pi(\omega,0)$ is not diagonal and the eigenvalues of its
spatial part are not all the same.


\subsection {Criteria for Instability}
\label {sec:criteria}

In momentum space $Q = (\omega,\q)$, the linearized effective equation
(\ref{eq:Maxwell3}) for the soft gauge
fields is
\begin {equation}
\label{eq:lin_eff_eq}
   [(-\omega^2 + q^2) g^{\mu\nu} - Q^\mu Q^\nu + \Pi^{\mu\nu}(\omega,\q)]
   A_\nu = 0 .
\end {equation}
For a given $\q$, there will be instabilities if there are any solutions
for $\omega$ with $\Im \omega > 0$.

For the case of parity-invariant background
particle distributions, $f(\p) = f(-\p)$,
there are some simple sufficient conditions for the existence of
instabilities.  In the rest of this paper,
{\em we will assume that distributions are parity invariant.}%
\footnote{
  We also implicitly assume that hard particle dispersion relations
  are parity invariant.
}
There is then a simple sufficient (but not necessary)
condition for the existence of
an instability:
\begin {quote}
   {\it Condition 1.}\/
   There is an instability associated with a given wavenumber $\q$
   for each negative eigenvalue of the $3\times3$ matrix
   $q^2 \delta^{ij} + \Pi^{ij}(0,\q)$.
\end {quote}
This is a simple generalization of the Penrose criteria used by
\Mrowczynski\ in Ref.\ \cite{mrow1} for a special case.
There is a simple continuity argument for this assertion, which we give
in Appendix \ref{app:condition1}.%
\footnote{
  Ref.\ \cite{mrow1} considered the special case where $\hat\q$ points
  along a symmetry axis of $f(\p)$
  and the eigen-modes of $\Pi^{ij}$ are determined by symmetry.
  Textbook discussions of the Penrose criteria for
  similar special cases in the non-relativistic context may be
  found in Sec.\ 9.10.2 of Ref.\ \cite{krall} and Sec.\ 3.3.4 of
  Ref.\ \cite{davidson}.
  We are unaware of a general analysis of the case where $\hat\q$ does not
  point in a symmetry direction of the problem,
  and so we give our own argument in the appendix.
  It also sets up the argument for Condition 2 below.
  Similarly, one often sees the Penrose criteria playing the role
  of {\it necessary}\/ (as well as sufficient) conditions for instability in
  certain situations, such as where $\hat\q$ is in the $z$ direction
  and $f(\p)$ has the form $F_z(p_z) F_\perp(p_\perp)$ with $F(p_z)$
  a monotonic decreasing function of $|p_z|$ \cite{mrow1,krall,davidson}.
  We do not know how to prove necessity in the case where
  $\hat\q$ points in non-symmetry
  directions, for reasons described in the appendix.
}

The zero-frequency self-energy $\Pi(0,\q)$ depends only on the direction
$\hat\q$ and not the magnitude $q$ of $\q$ because
(\ref{eq:pi}) can be rewritten for $\omega=0$ as%
\footnote{
  Physically, the independence of $\Pi(0,\q)$ on the magnitude $q$ of the
  soft field wave number depends on $q$
  being small compared to hard particle momenta---an
  approximation implicit in our use of the Vlasov equations.
}
\begin {equation}
   \Pi^{\mu\nu}(0,\q) =
   e^2 \int_\p
     \frac{\partial f(\p)}{\partial p^k}
     \left[ -v^\mu g^{k\nu}
            + \frac{v^\mu v^\nu \hat q^k}{\v\cdot\hat\q-i\epsilon}
     \right] .
\label {eq:pistatic}
\end {equation}
We will write $\Pi(0,\hat\q)$ instead of $\Pi(0,\q)$ to emphasize this.
A simple corollary of Condition~1 is then that
there will always be an instability in some mode
({\it i.e.}\ for small enough $q$)
if $\Pi^{ij}(0,\hat\q)$ has a negative eigenvalue.

\begin{figure}
\includegraphics[scale=0.70]{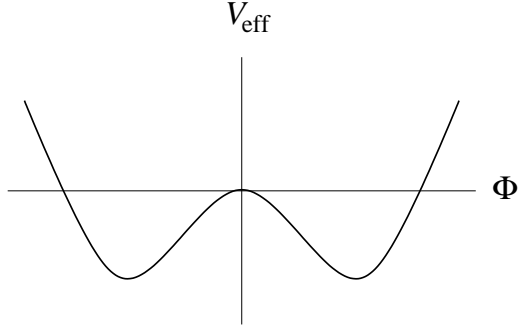}
\caption{%
    \label{fig:potential}
    Example of an effective potential $V_{\rm eff}(\Phi)$
    in scalar field theory for which $\Phi=0$ would be unstable.
    }
\end{figure}

Readers may find it useful to consider as an analogy the static
effective potential $V_{\rm eff}(\Phi)$ for the softest modes of a
scalar field theory.  If $V_{\rm eff}(\Phi)$ has negative curvature
at $\Phi=0$, as depicted in Fig.\ \ref{fig:potential}, then
$\Phi=0$ is unstable.  For scalars with no intrinsic mass (analogous
to gauge bosons), this curvature is simply $\Pi(0)$.  The condition
$\Pi(0) < 0$ implies that there is an instability, even though
the precise calculation of the growth rate would require studying
the dynamics, depending on $\Pi(\omega)$.  Linearizing
$\frac12|\grad\Phi|^2 + V_{\rm eff}(\Phi)$ about $\Phi=0$, one would
find that the condition for a particular mode to be unstable is
$q^2+\Pi(0,{\bm q}) < 0$, analogous to Condition 1.  The picture of
Fig.\ \ref{fig:potential} also suggests that non-linearities in the
interactions of the soft fields may eventually cut off the
growth of the instability, a topic we will return to later.

We will refer to instabilities discovered by Condition 1 as ``magnetic''
instabilities because the condition itself involves the self-energy for
$\A(\omega{=}0)$ and so involves magnetic but not electric fields.
We will give a physical picture of the origin of the instability in
Sec.\ \ref{sec:physical}.
Of course, the growth of magnetic fields implies the generation of
electric fields as well, by Maxwell's equations, so the actual growing
mode will not be purely magnetic.

Note that the Ward identity $q_\mu \Pi^{\mu\nu}=0$ implies
$\hat q^i \, \Pi^{ij}(0,\q)=0$.  Therefore, the longitudinal polarization
of $\A(\omega{=}0)$ is
always associated with a zero eigenvalue of $\Pi^{ij}(0,\q)$ and so
will not play a role in Condition~1.
We could have phrased Condition~1 equivalently using its transverse
projection:
\begin {quote}
   {\it Condition 1'.}\/
   There is an instability associated with a given wavenumber $\q$
   for each negative eigenvalue of the $3\times3$ matrix
   $q^2 \delta^{ij} -q^i q^j + \Pi^{ij}(0,\q)$.
\end {quote}

In Appendix \ref{app:anisotropy}, we expand upon an argument in
Ref.\ \cite{boltzmann} to use Condition 1 to show that some unstable mode
always exists unless the matrix $\Pi^{ij}(0,\hat q)$ is identically zero.
We further show that this leads to a simple necessary as
well as
sufficient condition for
magnetic instability in the ultra-relativistic limit:
\begin {quote}
   {\it Condition 1-b.}\/
   Magnetic instabilities exist for a given (parity symmetric)
   distribution $f(\p)$ if
   \begin {equation}
       {\cal M}(\hat\p) \equiv
       \frac{e^2}{2\pi^2} \int_0^\infty p^2 dp \> \frac{v}{p} \, f(p \hat\p)
   \label{eq:calM}
   \end {equation}
   is anisotropic in $\hat\p$.
   Moreover, in the ultra-relativistic limit, this is a necessary (as well
   as sufficient) condition for the existence of instabilities.
\end {quote}
We note in passing that the angular average of ${\cal M}(\hat\p)$ over
$\hat\p$ gives
\begin {equation}
   m_\infty^2
   \equiv \langle {\cal M}(\hat\p) \rangle_{\hat\p}
   = e^2 \int_\p \frac{v}{p} \, f(\p) ,
\label {eq:minfty}
\end {equation}
which defines a basic mass scale we will encounter repeatedly.
In the ultra-relativistic case, it also
turns out to be the effective mass of transverse electromagnetic waves
in the large $q$ limit, with $\omega^2 \simeq q^2 + m_\infty^2$.
This is shown in Ref.\ \cite{boltzmann} and inspires the notation
$m_\infty$, though this correspondence only holds in the
ultra-relativistic case.
Note that $m_\infty^2$ is independent of the direction $\hat\q$.

In Condition 1-b we have re-emphasized the underlying assumption of our
discussion that $f(\p)$ is parity symmetric.  The condition is obviously
incorrect without this assumption: Consider, for example, a standard
equilibrium
distribution boosted to an inertial frame where the average momentum
does not vanish.
The corresponding $f(\p)$ and ${\cal M}(\p)$
are then anisotropic and also not parity invariant, but equilibrium
is stable, regardless of the inertial frame.

There is another type of instability that can occur, which can be
associated with the electric potential $A^0$ at $\omega=0$
[which for non-zero $\omega$ is related to the longitudinal polarization of
$\A$ by the Ward identity (\ref{eq:ward})].
In Appendix \ref{app:condition2}, we show that another (independent)
sufficient condition
for the existence of an instability is
\begin {quote}
   {\it Condition 2.}\/
   There is an instability associated with a given wavenumber $\q$
   if
   \vspace{-2pt}
   \begin {equation}
      q^2 - \Pi^{00}(0,\hat\q)
      + \Pi^{0i}(0,\hat\q) \left[q^2+{\bm\Pi}(0,\q)\right]^{-1}_{ij}
        \Pi^{j0}(0,\hat\q)
      < 0 ,
   \label {eq:condition2}%
   \vspace{-2pt}
   \end {equation}%
   where $[q^2+{\bm\Pi}]^{-1}$ denotes the inverse of the
   $3\times3$ matrix
   $q^2 \delta^{ij}+\Pi^{ij}$.
\end {quote}
In the non-relativistic limit, $\Pi^{i0}$ is suppressed by powers of
$v/c$ compared to $\Pi^{00}$ and this condition becomes simply
$q^2 - \Pi^{00}(0,\hat\q) < 0$, which will happen for some $\q$
if $- \Pi^{00}(0,\hat\q) < 0$.
[A simple mnemonic for this condition
is to recall that, for isotropic distributions,
$-\Pi^{00}(0,\q)$ gives the squared Debye mass $\md^2$.  So
$-\Pi^{00}(0,\hat q) < 0$ is analogous to $\md^2$ having the wrong sign,
which one might guess could indicate instability.]
This electrostatic instability is routinely discussed in
plasma physics texts%
\footnote{
   For example, a small sampling of references we have found useful is
   Refs.\ \cite{krall,davidson,goldston}.
}
and is typically called the two-stream or Buneman
\cite{buneman}
instability.
We will refer to instabilities discovered by Condition 2 as
``electric'' instabilities, even though the situation is a little
more complicated in the relativistic case.

Because $\hat q^i \Pi^{0i}(0,\hat q) = 0$, the $\hat\q$ polarization
does not contribute in the last term of (\ref{eq:condition2}). 
One may then restrict the sums over $i$ and $j$ to directions transverse
to $\q$ and similarly project ${\bm \Pi}$ to a $2\times2$ matrix in
the transverse subspace.

Were it not for the $i\epsilon$ prescription in the retarded
self-energy (\ref{eq:pi}), $\Pi^{0i}(0,\q)$ would vanish by
parity ($\p \to -\p$).  Using
$(x-i\epsilon)^{-1} = {\rm P.P.}~x^{-1} + i\pi\,\delta(x)$, where
${\rm P.P.}$ denotes principal part, parity then gives
\begin {equation}
   \Pi^{0i}(0,\q) =
   i\pi e^2 \int_\p
     v^i \, \delta(\v\cdot\q) \, \q\cdot\grad_\p f(\p) ,
\label{eq:Pi0i}
\end {equation}
which is pure imaginary.  We may therefore rewrite Condition 2
as
\begin {equation}
      q^2 - \Pi^{00}(0,\hat\q)
      - [\Im \Pi^{0i}(0,\hat\q)] \left[q^2+{\bm\Pi}(0,\hat\q)\right]^{-1}_{ij}
        [\Im \Pi^{j0}(0,\hat\q)]
      < 0 \, .
\end {equation}
If there are no magnetic instabilities, so that $q^2+{\bm\Pi}(0,\q)$ is
positive definite, then the last term contributes negatively to the
left-hand side.  In this case, a {\it weaker}\/ sufficient condition
for an electric
instability is then that $q^2 - \Pi^{00}(0,\hat\q) < 0$.

Our categorization of instabilities as ``magnetic'' or ``electric''
is not always physically significant.  We use this terminology to
tell us whether the existence of the instability was indicated by
Condition 1 or Condition 2 respectively.  Condition 1 is related to
the transverse eigen-values of the zero-frequency self-energy
$\Pi^{ij}(0,q)$, and Condition 2 is related to the $\hat\q$ polarization
[see Appendix \ref{app:condition2}].  However, the actual growing unstable
solutions to the dispersion relation (\ref{eq:lin_eff_eq}) have
{\it non}-zero frequency $\omega$ (with $\Im\omega>0$).
The corresponding polarizations can look quite
different from those of $\Pi(0,\q)$
in generic situations where
eigen-polarizations of $\Pi^{ij}(\omega,\q)$
are not fixed by symmetry.  In particular, we will see a specific
example in Sec.\ \ref{sec:planar} where, for certain limiting cases of $\q$,
the actual growing unstable modes for a
``magnetic'' instability and ``electric'' instability end up being
the same up to a rotation about an axis of symmetry.
In the qualitative discussion that follows, however, we will focus on
the Penrose conditions themselves and the nature of the self-energy at
zero frequency.


\subsection {Qualitative Origin of Instabilities}
\label{sec:physical}

We will now give a brief qualitative review of the origin of the
instabilities we have discussed, in part because we do not know of a
really good qualitative review in the non-relativistic plasma literature
(especially of how the instability is saturated in the Abelian case),
and in part to
emphasize in a few cases differences for ultra-relativistic plasmas.
In the context of the QCD plasma literature, a brief sketch is
given by \Mrowczynski\ in Ref.\ \cite{mrow3}, which we will expand
upon.  Understanding how the instabilities are generated is useful
for intuition in understanding what types of instabilities may arise
for a given $f(\p)$.

We will start with a formal observation about Condition 1 for
Weibel instabilities.  It is useful to integrate the formula
(\ref{eq:pi}) by parts to obtain%
\footnote{
   Here and henceforth, we implicitly assume free hard particle dispersion
   relations are isotropic so that the direction of $\v$ is the same as
   that of $\p$.  Kinetic theory is sometimes
   used to describe the effects of hard modes in classical thermal
   field theories on a discrete lattice, where this assumption would not
   be true.
}
\begin {multline}
   \Pi^{ij}(\omega,\q)
   = e^2 \int_\p f(\p) \biggr\{
       \frac{v}{p} \left[
          \delta^{ij}
          - \frac{q^i v^j+q^j v^i}{-\omega+\v\cdot\q-i\epsilon}
          + \frac{(-\omega^2+q^2) v^i v^j}{(-\omega+\v\cdot\q-i\epsilon)^2}
         \right]
\\
       + \frac{dv}{dp} \,
           \frac{\omega^2 v^i v^j}{(-\omega+\v\cdot\q-i\epsilon)^2}
     \biggl\}.
\label {eq:pi2}
\end {multline}
The last term vanishes for ultra-relativistic theories, for which
$v=1$ and $dv/dp = 0$.
The interesting property of this expression is that, if one were to ignore
the $i\epsilon$ prescription, the integrand above would be positive
semi-definite for $\omega=0$.  If not for the
contribution near the singularity,
$\Pi^{ij}(0,\hat\q)$ could never have a negative eigenvalue.
To see this, consider any unit polarization vector $\bpol$
and use (\ref{eq:pi2}) to write
\begin {equation}
   \pol^i \, \Pi^{ij}(0,\q) \, \pol^j
   = e^2 \int_\p \frac{v}{p} \, f(\p) \,
       \frac{\left[\pol^2(\v\cdot\q)^2
             - 2(\bpol\cdot\q)(\bpol\cdot\v)(\v\cdot\q)
             + q^2(\bpol\cdot\v)^2\right]}
            {(\v\cdot\q-i\epsilon)^2} ,
\label {eq:pol}
\end {equation}
whose integrand, if not for the $i\epsilon$ prescription, would always be
non-negative by the
vector inequality
\begin {equation}
   a^2 (\b\cdot\c)^2 - 2(\a\cdot\b)(\b\cdot\c)(\c\cdot\a) + b^2 (\a\cdot\c)^2
   \ge 0 .
\end {equation}

The conclusion is that, in the perturbative limit,
the origin of the Weibel instability must come completely from
hard particles with $\v$ perpendicular to $\q$, which is when the
$i\epsilon$ prescription plays a role.
All other hard particles are stabilizing.
As we shall review, the destabilizing particles with $\v$ perpendicular
to $\q$ are just the perturbative limit of particles whose motions
in the $\q$ direction
are trapped by the fields.
The importance of trapped particles to discussions of instabilities,
and their relation to singularities of integrals in perturbative calculations,
has a long history starting with the somewhat analogous case of
electric instabilities in one dimension \cite{bohm&gross,bgk}.
Note also that the effect of the stabilizing
particles in (\ref{eq:pol})
generically diverges as $\cos\theta\to0$ like
$d(\cos\theta)/\cos^2\theta$,
where $\theta$ is the angle between ${\bm p}$ and ${\bm q}$.

\begin{figure}
\includegraphics[scale=0.40]{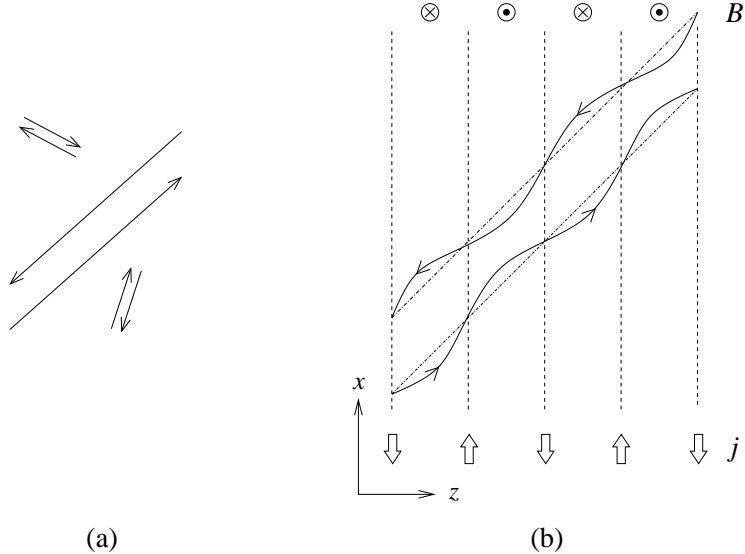}
\caption{%
    (a) Parity opposites in a gas of free-streaming particles.
    (b) Small deviations from straight-line trajectories of untrapped
        (stabilizing) particles in the background
        of a small magnetic field $\B = {\cal B} {\bf e}_y \sin(q z)$.
        The arrows show the direction of motion of positively charged
        particles, or the direction opposite to that of negatively charged
        particles.  The direction $(\pm y)$
        of the magnetic field is shown at the top, and the vertical
        dashed lines correspond to $\B(z)=0$.
        The direction ($\pm x)$
        of the average current $\j(z)$ caused by the particles
        is shown at the bottom.
    \label{fig:mag1}
    }
\end{figure}

To understand all this qualitatively, consider first a gas of non-interacting
hard particles in zero field, as shown in Fig.\ \ref{fig:mag1}a.
By our parity assumption, there is no current $\j(\x)$ associated with
these particles because, for all the particles going in one direction,
there are just as many particles going in the other direction.
Now turn on a small magnetic field $\B$ with wave-vector $\q$,
which
we shall take to point in the $z$ direction.
Take $\B$ to be in the $\pm y$ direction with
$\B = {\cal B} {\bf e}_y \sin(q z)$, as depicted in Fig.\ \ref{fig:mag1}b,
where ${\bm e}_y$
is the unit vector in the $y$ direction.
We'll take the vector field $\A$ to be in the $\pm x$ direction, with
$\A = {\cal A} {\bf e}_x \cos(qz)$ and $\B = \grad\times\A$.
Magnetic forces from the small $\B$ field will make charged particles
slightly wiggle
around straight-line trajectories.  The small wiggles in
direction will cause the $x$-component $j_x$ of the current to 
be larger in some places and smaller in others, compared to the value
obtained from the straight-line trajectory.  This is
shown at the bottom of the figure.
These currents set up a magnetic field that opposes
the original magnetic field and so stabilize against its growth.

The radius of curvature of particle trajectories in the plane
perpendicular to the magnetic field is
\begin {equation}
   R = \frac{p_\perp}{eB} \,,
\label{eq:R}
\end {equation}
where $p_\perp$ is the momentum in that plane.
It's easy to estimate%
\footnote{
  The field has roughly the same order of magnitude over a range of $z$
  close to half a period $\lambda/2=\pi/q$ of the magnetic field.
  Simple geometry gives that the resulting small change in $v_x$
  over half a period is then of order $v\lambda/R$, which in turn is
  $\sim e v B/pq \sim e v A/p$.
}
that $\Delta v_x \sim e A v/p$ so that the
effect described gives a contribution to $j_x$ of order
$e^2 A \int_\p f v/p$.  This contribution to the current is finite in
the $\cos\theta \to0$ limit and does not explain the small
$\cos\theta$ divergence
discussed earlier.

Besides this variation in the {\em magnitude} of $j_x$ for each particle
at a given position, there is a
second effect: the amount of {\em time} the particle spends
at a given $z$ coordinate changes, because the component $v_z$ of its
velocity also wiggles.
The particles will spend more of their time in
regions of $z$ where $v_z$ is smaller (which are the regions where
$v_x$ is bigger).  The slowing/speeding effect of the $z$ motion
therefore generates a contribution to the average
current $\j(z)$ which is also in the directions
shown in Fig.\ \ref{fig:mag1}b.  One may parametrically estimate
that the variation $\Delta v_z$ of $v_z$ has relative size
of order%
\footnote{
   Motion in a magnetic field conserves $|\v|$.  This means that
   $v_x \, \Delta v_x + v_z \, \Delta v_z \simeq 0$,
   so that $(\Delta v_z)/v_z \sim (\Delta v_x) v_x/v_z^2$.  Then, use
   the previous estimate of $\Delta v_x$.
}
$(\Delta v_z)/v_z \sim e A v v_x/(p v_z^2)$,
giving
\begin {equation}
   j^x \sim e^2 A \int_\p f \, \frac{v v_x^2}{p v_z^2} .
\label{eq:jx}
\end {equation}
This gives the $d(\cos\theta)/\cos^2\theta$ divergence, since
$v_z = v \cos\theta$ in our choice of coordinates.
This effect is also stabilizing.

\begin{figure}
\includegraphics[scale=0.40]{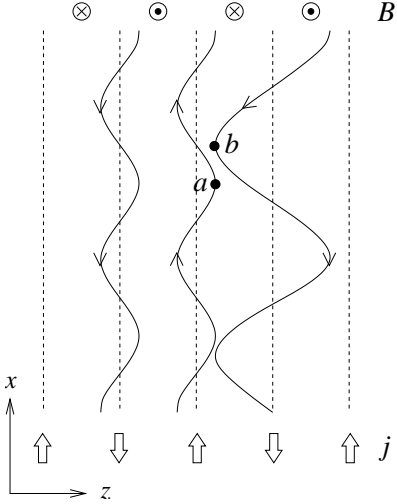}
\caption{%
    The same situation as in Fig.\ \ref{fig:mag1}b but for particles
    with small enough $v_z$ that their $z$ motion is trapped.
    \label{fig:mag2}
    }
\end{figure}

So far we have focused on the contribution from particles that stabilize
magnetic fluctuations.  Now consider particles whose initial velocities
are very close to being orthogonal to $\q$.  The curvature of trajectories
caused by the magnetic field will trap the $z$ motion of such
particles, as depicted in Fig.\ \ref{fig:mag2}.
The two trajectories
shown on the left of the figure generate currents {\it opposite}\ those
shown in Fig.\ \ref{fig:mag1}, which therefore generate a magnetic field
that adds to the existing one and contributes to instability.
The contribution of the trajectory on the far right can be appreciated
as follows.  Imagine initially two particles, one at the point marked
{\it a}\/
and one at the point marked {\it b}, have parity-opposite momentum in the
$\pm x$ directions.  Then imagine suddenly turning on the magnetic forces
in the problem.  The upward moving particle at {\it a} will follow the middle
trajectory and continue to give an upward contribution to
$\j$ in roughly the same region of $z$.  The downward particle at {\it b},
whose contribution to $\j$ was originally canceling, will drift far away
and so spend only a portion of its time
close to its original $z$.  That leads to the generation of a new
upward average current in the region of their original $z$.

For very small magnetic fields, only a perturbatively small
fraction of phase space,
corresponding to very small $|v_z|$, is trapped.  However, each such
particle contributes to the current separation an $O(e)$ amount, {\em
unsuppressed} by the size of $A$.  In contrast, the untrapped particles,
making up almost all of phase space, each contribute a stabilizing
influence; but the size of that stabilizing influence is 
perturbatively small (that is, suppressed by a power of the size of
$A$).  It is for this reason that trapped and
untrapped contributions can be the same order of magnitude even in the
perturbative limit.%
\footnote{
  Details depend on the distribution $f(\p)$ and the direction $\hat\q$.
  However, for generic distributions and generic directions $\hat\q$,
  one can parametrically estimate that trapping typically occurs in weak
  fields when $R|\cos\theta| \lesssim 1$.  From (\ref{eq:R}), that is
  $|\cos\theta| \lesssim \sqrt{eA/p}$, which indeed is formally
  arbitrarily small in the perturbative limit $e \to 0$.
  In a hand-waving sense, the $i\epsilon$ prescription in the
  denominator of (\ref{eq:pol}) can then be thought of as being of
  order $\sqrt{eA/p}$.
  In any case, one can use (\ref{eq:jx}) and integrate down to this
  $|\cos\theta|$
  to parametrically estimate the stabilizing untrapped contribution to
  be of order $e \int_\p f v /\sqrt{e A/p}$.  One finds a similar
  estimate for the order of magnitude of the trapped contribution,
  and the leading $O(e^{1/2})$ pieces of these contributions cancel
  to leave the
  perturbative $O(e^2)$ result for $\Pi^{ij}(0,\q)$.
}

\begin{figure}
\includegraphics[scale=0.40]{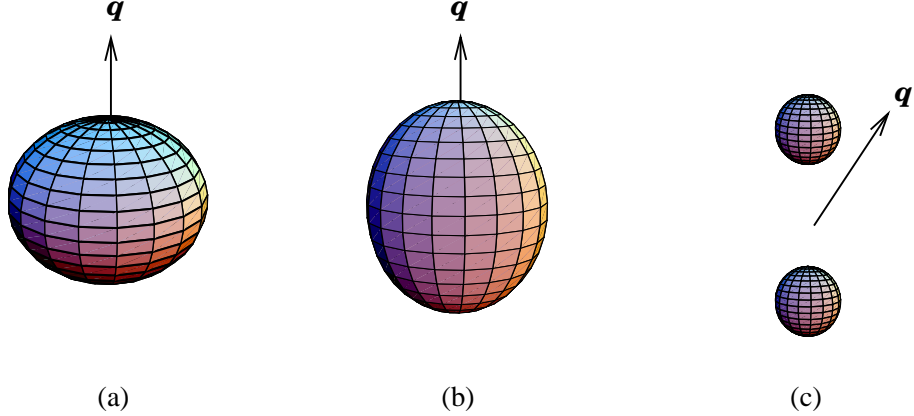}
\caption{%
    Some qualitatively different examples of anisotropic distributions
    $f(\p)$ together with some specific examples of directions for $\q$.
    The plots show the shape of surfaces of constant $f$ in $\p$ space.
    \label{fig:squash}
    }
\end{figure}

In the case of isotropic hard distributions $f(\p)$, the contributions
of trapped (destabilizing) and untrapped (stabilizing) particles must
cancel to give the isotropic result $\Pi^{ij}(0,\hat\q)=0$.
This delicate cancellation
allows us to understand the presence or absence of magnetic
instabilities in certain simple situations.
If we start with an isotropic distribution and, for a given $\q$, remove some
of the untrapped particles, then we must have magnetic instability.
This is the
case for the $\q$ shown for an oblate distribution in
Fig.\ \ref{fig:squash}a.  If we instead add untrapped particles, we must
have stability, such as for the $\q$ shown for a prolate distribution in
Fig.\ \ref{fig:squash}b.  In \ref{fig:squash}c, assume that the number
of particles outside the surfaces is insignificant.  Then
this is another example where there is no magnetic instability (for the
$\q$ shown) because the number of trapped particles is insignificant.

\begin{figure}
\includegraphics[scale=0.40]{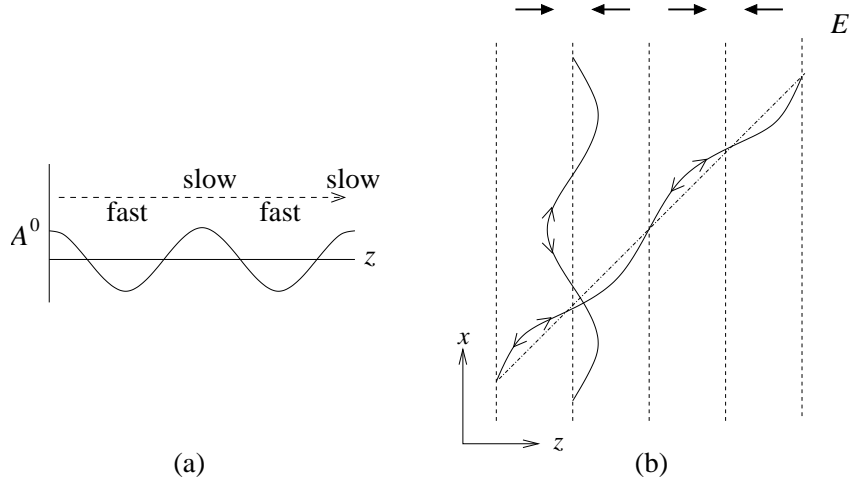}
\caption{%
    Qualitative picture of particle motion in a sinusoidal electric
    potential.
    (a) An untrapped non-relativistic particle in one dimension.
    (b) Untrapped and trapped particles in two or three dimensions.
    \label{fig:elec}
    }
\end{figure}

Finally, we will review the related picture for electric instabilities.
We will focus on $\Pi^{00}(0,\hat\q)$, which gives the charge response to
a sinusoidal electric potential.  Consider a small static electric
potential $A^0 = \phi \cos(q z)$ and the corresponding electric
field $\E = -\grad A^0 = {\cal E} {\bm e}_z \sin(q z)$.  First think
about the one dimensional problem---that is, particles moving in the
$z$ direction, depicted in Fig.\ \ref{fig:elec}.  If the field is small,
typical particles will be able to move over the electric potential barriers.
By energy conservation, however, positively charged particles will move
slower at the maxima of $A^0$ and faster at the minima.  So they will
spend more of their time near the maxima, and so their contribution to
the average charge density will be greatest there, which will generate a
yet larger $A^0$ there.  (Negatively charged particles will spend more
time at the minima of $A^0$, which will also enhance the magnitude of $A^0$.)
Untrapped particles are therefore destabilizing, which is the opposite of
what happened in the magnetic case.  In contrast, positively charged
particles which have tiny velocities are trapped in the minima of $A^0$
and so contribute positively to the charge density there, which reduces
$A^0$.  The electrically trapped particles are stabilizing.
The one-dimensional situation is often discussed in the non-relativistic
case but is not relevant to ultra-relativistic
particles because such particles always move with $v=1$.  However, in the
three-dimensional case, $v_z$ can become smaller or larger as particles
move through minima and maxima even while $v$ remains the same, just
as in our discussion of the magnetic case.
Examples of electrically trapped and untrapped particles are given
in Fig.\ \ref{fig:elec}b.

Isotropic distributions are electrically stable and associated with
a positive value of $-\Pi^{00}(0,\q)$ given by the squared Debye mass
$\md^2$.  Small deviations from isotropy will not satisfy Condition 2
of Sec.\ \ref{sec:criteria}, but significant deviations from
isotropy can.

An example where one can simply conclude that Condition 2 for electric
instability is not satisfied is an oblate distribution, such as shown
in Fig.\ \ref{fig:squash}a, with $\hat\q$ pointing along the axis of
axial symmetry, which we will take to be the $z$ axis.
For reasons similar to those given previously for the magnetic case,
except that now the role of trapped and untrapped particles is
reversed, $-\Pi^{00}(0,{\bm e}_z) > 0$.
Axial symmetry and the Ward identity (\ref{eq:ward}) imply
$\Pi^{0i}(0,{\bm e}_z)$ vanishes,
and so Condition 2 will not be satisfied in this situation.


\subsection {Saturation of Instabilities}

When discussing the linear response of the gauge fields, there is no
significant difference between Abelian and non-Abelian theories.
When the instabilities grow large enough to become non-linear, there is.
It is useful to understand qualitatively how instabilities
saturate in the Abelian case, in order to contrast it with the
non-Abelian case.

The saturation of Abelian electric instabilities is straightforward
to understand qualitatively.  Return to Fig.\ \ref{fig:elec} and
imagine making the electric field arbitrarily large (holding
$q$ fixed).  Then essentially all particles will become trapped
in the $\q$ direction
near the bottom of the potential wells, which means there will
be no instability.  So, if there was originally an instability, the
field will eventually grow large enough for it to saturate.
This will happen very roughly when the electric potential energy
$e A^0$ becomes of order the typical energy associated with the
motion of the particles in the $\q$ direction.
In the non-relativistic limit, that is
\begin {subequations}
\label {eq:Esaturate}
\begin {equation}
   A^0 \sim \frac{p_\parallel^2}{e M} \, ,
   \qquad
   E \sim \frac{p_\parallel^2 q}{e M} \, ,
\end {equation}
where $M$ is the particle mass and 
$p_\parallel$ indicates the typical value of the component
of $p$ parallel to $\q$.
In the ultra-relativistic limit,
\begin {equation}
   A^0 \sim \frac{p_\parallel^2}{e p} \, ,
   \qquad
   E \sim \frac{p_\parallel^2 q}{e p} \, .
\end {equation}
\end {subequations}
This saturation was used long ago to construct non-linear wave solutions
in the non-relativistic Abelian case, known as Bernstein-Greene-Kruskal waves
\cite{bgk}.

\begin{figure}
\includegraphics[scale=0.40]{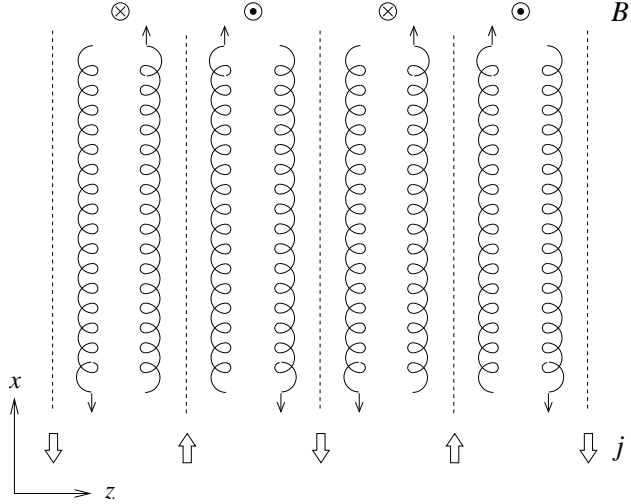}
\caption{%
    Trajectories in an arbitrarily large magnetic field.  The size of the
    orbits and the rate of their drift has been exaggerated.
    \label{fig:mag3}
    }
\end{figure}

To qualitatively understand the saturation of magnetic instabilities,
return to the pictures of Figs. \ref{fig:mag1}
and \ref{fig:mag2}, but now make the magnetic fields arbitrarily large
(holding $\q$ fixed).
In this case, the typical radius of curvature (\ref{eq:R}) will be
small compared to the distance in $z$ over which $\B(z)$ changes sign.
Instead of looking like Fig.\ \ref{fig:mag1}b or \ref{fig:mag2}, typical
trajectories will instead look like Fig.\ \ref{fig:mag3}.  The particles
follow small, nearly circular orbits in the magnetic field which slowly
drift in the direction indicated due to slightly higher curvature on the
side where $\B$ is slightly larger.  This sets up currents, shown at
the bottom of the figure, which create magnetic fields that oppose
the original field.  Essentially all particles are therefore stabilizing
in the high field limit.  So, once again,
if there was originally an instability, the
fields will eventually grow large enough for it to saturate.  That will
happen very roughly when the radius of curvature (\ref{eq:R}) becomes of
order the wavelength $\lambda \sim 1/q$ of the field, so%
\footnote{
   This and (\ref{eq:Esaturate}) represent estimates in generic cases
   of significant anisotropy.  If, for example, the anisotropy is very tiny
   or $q$ is very close to its maximum unstable value, then
   a very tiny higher-order correction to the perturbative calculation might
   change instability to stability.
   A brief and somewhat different discussion of rough estimates of the
   saturation condition in the non-relativistic case for both magnetic and
   electric instabilities may be found in
   Ref.\ \cite{davidson&etal}, as well as references therein.
   There the saturation condition is estimated in terms of time rather
   than distance scales by equating the ``bounce'' time for an oscillation
   of a trapped particle to the linearized growth rate of the instability.
   For generic situations and generic unstable $\q$, this is roughly
   equivalent to our rough conditions (\ref{eq:Esaturate}) and
   (\ref{eq:saturateAbelian})
   if one assumes that the growth rate $\gamma$ is of order $v q$ (where we
   will no longer try to distinguish between $v_\parallel$ and $v_\perp$).
   See, for example, Eqs.\ (1), with the condition
   $\omega_{\rm B} \sim \gamma$,
   and (114), with $k$ small, of Ref.\ \cite{davidson&etal}.
   The estimate $\gamma \sim v q$ (which is $\ll q$ in the non-relativistic
   limit) can be roughly understood
   by thinking of increasing $\omega = i\gamma$ from zero and noting
   that it first has an effect in the self-energy
   when $\omega \sim vq$, so that it significantly affects the integration
   over the denominators in (\ref{eq:pi2}).
}
\begin {equation}
   B \sim \frac{p_\top^{} \, q}{e} \,,
   \qquad
   A \sim \frac{p_\top^{}}{e} \,.
\label {eq:saturateAbelian}
\end {equation}
A simple mnemonic for cases where the hard particles can be
described as excitations in a quantum field theory is to realize that,
when acting on hard excitations of momentum $p$,
the gauge term in the covariant derivative $D = \partial - i e A$
cannot be treated as a small perturbation to the derivative term when
(roughly speaking) $A \gtrsim p/e$.

In contrast, if different unstable soft field modes with momenta of order $q$
can interact with each other, then linearity
breaks down when
\begin {equation}
   A \sim \frac{q}{g}
\label{eq:soft_saturate}
\end {equation}
instead of $A \sim p/g$,
where we have switched notation from $e$ to $g$ as a way of emphasizing that
this is special to the non-Abelian case.
We will see a specific example in Sec.\ \ref{sec:twirl}.

Even for the Abelian case, the picture of Fig.\ \ref{fig:mag3} is a
bit simplistic.  It assumes
that only one unstable mode with a given $\q$ has been excited.
In most situations, an entire spectrum of unstable modes will grow
together and then interact with each other once they become
non-perturbatively large according to (\ref{eq:saturateAbelian}).
The idealized case of a single Abelian mode has been
studied in the non-relativistic
literature and certain non-linear wave solutions
have been found, known as magnetic
Bernstein-Greene-Kruskal (BGK) waves \cite{berger&davidson,davidson&etal}.
The generation of these non-linear waves has
also been investigated numerically
\cite{davidson&etal,yang2,califano},
in some cases with only
a single unstable mode significantly excited and in other cases with many.


\section {Instability of the planar momentum distribution}
\label{sec:planar}

We will now
focus in detail on the planar momentum distribution of
(\ref{eq:fplanar}):
\begin {equation}
   f(\p,\x) = F(p_\perp) \, \delta(p_z) .
\end {equation}


\subsection {Magnetic instabilities}

\subsubsection {\boldmath$\omega=0$ magnetic stability analysis}

\begin{figure}
\includegraphics[scale=0.40]{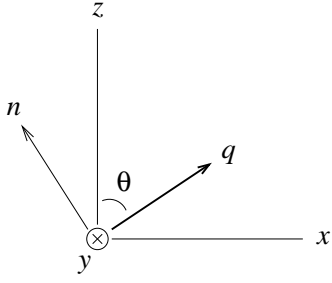}
\caption{%
    Convention for labeling directions with respect to $\hat\q$, which is
    taken to lie in the $xz$ plane.
    \label{fig:directions}
    }
\end{figure}

To check for evidence of magnetic instabilities,
first consider the self-energy (\ref{eq:pi2}) at $\omega=0$,
\begin {equation}
   \Pi^{ij}(0,\hat\q) =
   e^2 \int_\p f(\p) \, \frac{v}{p} \, \left[
          \delta^{ij}
          - \frac{\hat q^i \hat p^j+\hat q^j \hat p^i}
                 {\hat\p\cdot\hat\q-i\epsilon}
          + \frac{\hat p^i \hat p^j}
                 {(\hat\p\cdot\hat\q-i\epsilon)^2}
   \right] .
\end {equation}
The planar distribution
$f(\p)$ has support only in the $p_x p_y$ plane and is
independent of the direction of $\p$ in that plane.
We may therefore factor
out the angular dependence of the integral and  write
\begin {equation}
   \Pi^{ij}(0,\q)
   = m_\infty^2 \left\langle
          \delta^{ij}
          - \frac{\hat q^i \hat p^j+\hat q^j \hat p^i}
                 {\hat\p\cdot\hat\q-i\epsilon}
          + \frac{\hat p^i \hat p^j}
                 {(\hat\p\cdot\hat\q-i\epsilon)^2}
         \right\rangle_{\hat\p\mathop{\in}\mbox{plane}} \, .
\label {eq:piplanar}
\end {equation}
Because of the axi-symmetry of the problem, we can assume without
loss of generality that $\q$ points in the $xz$ plane.
Then ${\bm e}_y$ is an eigenvector of $\Pi^{ij}$ because of
$y \to -y$ reflection symmetry.
For $\omega=0$, the direction $\hat\q$ is also an eigenvector of
$\Pi^{ij}$, corresponding to zero eigenvalue.  Let $\hat\n$ be the
orthogonal direction in the $xz$ plane, as depicted in
Fig.\ \ref{fig:directions}.
Let $\theta$ be the angle
$\hat\q$ makes with the $z$ axis and $\phi$ the angle $\hat\p$ makes with
the $x$ axis.  Then, for $\sin\theta\not=0$,
the eigenvalues of $\Pi^{ij}(0,\hat\q)$ are easily
computed from (\ref{eq:piplanar}) to be
\begin {subequations}
\label {eq:Piresult}
\begin {eqnarray}
   \Pi^{\hat\q\hat\q}(0,\hat\q) &=& 0 ,
\\
   \Pi^{\hat\n\hat\n}(0,\hat\q) &=&
   m_\infty^2 \int_0^{2\pi} \frac{d\phi}{2\pi}
      \left[ 1 + 0 + \frac{\cos^2\theta}{\sin^2\theta} \right]
   = \frac{m_\infty^2}{\sin^2\theta}
\label{eq:Pinnstatic}
\\
   \Pi^{yy}(0,\hat\q) &=&
   m_\infty^2 \int_0^{2\pi} \frac{d\phi}{2\pi}
      \left[ 1 + 0 + \frac{\sin^2\phi}{\sin^2\theta(\cos\phi-i\epsilon)^2}
      \right]
   = m_\infty^2 \left(1-\frac{1}{\sin^2\theta}\right)
   = - m_\infty^2 \cot^2\theta .
\label {eq:piyy}
\nonumber\\
\end {eqnarray}
\end {subequations}
Only $\Pi^{yy}(0,\hat\q)$ is negative.%
\footnote{
   This instability corresponds to poles in $\Delta_A$ of Ref.\
   \cite{strickland} in the $\xi\to+\infty$ limit.
   The possibility of instability in this mode was
   overlooked in Ref.\ \cite{randrup&mrow}.
}
By Condition 1 of
Sec.\ \ref{sec:criteria}, there
is a magnetic instability associated with $\A$ polarized in the
$y$ direction (and so $\B$ in the $\hat\n$ direction) whenever
\begin {equation}
   q < q_{\rm max}(\theta) \equiv m_\infty \cot\theta .
\label {eq:qmax}
\end {equation}

It is easy to understand qualitatively why there is no magnetic instability
associated with the $\n$ polarization of $\A$,
which would correspond to the magnetic
field $\B$ in the $\pm y$ direction.  Magnetic instability is caused by
trapped particles, which perturbatively are those with $\p$ orthogonal
to $\q$.  For the planar distribution with
$\q$ in the $xz$ plane, the only trapped particles are those with
$\p$ pointing in the $\pm y$ direction (for $\sin\theta\not=0$).
If $\B$ is also in the $\pm y$
direction, then the motion of these particles is not affected by the
magnetic field, and so there is no destabilizing contribution.


\subsubsection {Magnetic instability growth rate in the 
	ultra-relativistic limit}

\begin{figure}
\includegraphics[scale=0.40]{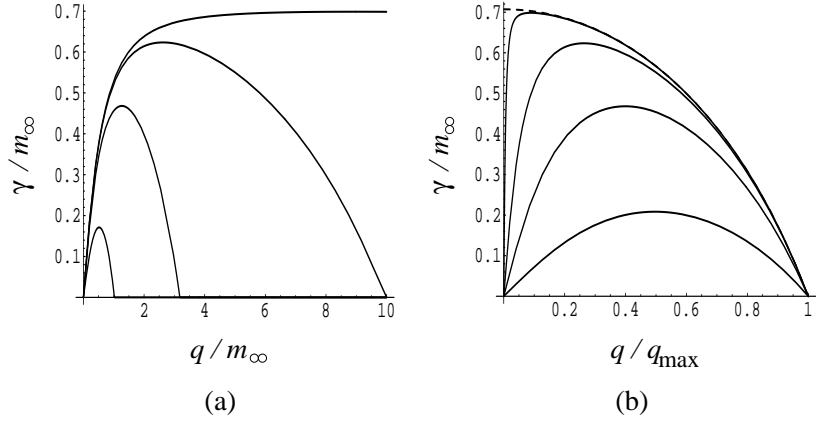}
\caption{%
    The growth rate vs.\ $q$ for magnetic instabilities for the planar
    distribution for various values of $\theta$.
    {}From the top down, the values are $\sin\theta$ = 0.01, 0.1, 0.3, 0.65.
    The small dashed line at the top of (b) shows the $\sin\theta\to0$ limit.
    \label{fig:results}
    }
\end{figure}

We can get simple expressions for $\Pi^{ij}(\omega,\q)$ for
general $\omega$ if we
specialize to the ultra-relativistic limit, which we will now do.
In this case, it is straightforward to
keep the frequency dependence in the manipulations that led to
from (\ref{eq:pi2}) to (\ref{eq:piplanar}) for the self-energy,
giving
\begin {equation}
   \Pi^{ij}(\omega,\q)
   = m_\infty^2 \left\langle
          \delta^{ij}
          - \frac{\hat q^i \hat p^j+\hat q^j \hat p^i}
                 {-\eta+\hat\p\cdot\hat\q-i\epsilon}
          + \frac{(-\eta^2+1) \hat p^i \hat p^j}
                 {(-\eta+\hat\p\cdot\hat\q-i\epsilon)^2}
         \right\rangle_{\hat\p\mathop{\in}\mbox{plane}} \, ,
\label {eq:piplanarA}
\end {equation}
where
\begin {equation}
   \eta \equiv \frac{\omega}{q} \, .
\end {equation}
Since there is no suggestion of magnetic instabilities for other
polarizations for $\sin\theta \not=0$, let us focus on $\Pi^{yy}$.
The nice thing
about the simple planar distribution is that one can find
a simple closed form solution from (\ref{eq:piplanarA}).  Performing
the $\phi$ integrals, one finds
\begin {subequations}
\label {eq:dispyy}
\begin {equation}
   \Pi^{yy}(\omega,\q) =
   m_\infty^2 \left\{
      1 +
      \frac{1-\eta^2}{\sin^2\theta}\left[\left(1-
           \frac{\sin^2\theta}{\eta^2}\right)^{-1/2} - 1 \right]
   \right\} .
\end {equation}
The corresponding dispersion relation is
\begin {equation}
   (-\eta^2+1)q^2 + \Pi^{yy}(\omega,\q) = 0 .
\label{eq:disp}
\end {equation}
\end {subequations}
By re-arranging terms algebraically and squaring, one can convert this
into a cubic equation in $\eta^2$, a subset of whose roots are the solutions
to (\ref{eq:disp}).  However,
the explicit closed form of that solution is no more
enlightening than a numerical solution.
In Fig.\ \ref{fig:results}a, we show the result for the growth rate
$\gamma = \Im\omega = q\Im\eta$ of the unstable solution
as a function of $q$ for various values of $\theta$.
It is clear that the largest growth rates correspond to $\sin\theta \to 0$,
which is not surprising since, for $\theta=0$, all particles in the planar
distribution become trapped particles, contributing to destabilization.
Plotting $\gamma/m_\infty$ vs.\ $q/q_{\rm max}$ in
Fig.\ \ref{fig:results}b, with $q_{\rm max}$ taken from (\ref{eq:qmax}),
it is clear that there is a limiting behavior as
$\sin\theta \to 0$.  For small $\sin\theta$,
the dispersion relation (\ref{eq:disp}) can be more compactly solved
to give
\begin {equation}
   \frac{\gamma}{m_\infty} \simeq
   f_1\left(\frac{q}{m_\infty}\right)
   + f_2\left(\frac{q}{q_{\rm max}}\right)
   - \frac{1}{\sqrt2} ,
\label{eq:Bsmall1}
\end {equation}
where
\begin {eqnarray}
   f_1(z) &=& \half \left\{ \bigl[(1+2z^2)^2+8z^2\bigr]^{1/2}
                          - (1+2z^2)\right\}^{1/2} ,
\label {eq:f1z}
\\
   f_2(z) &=& \frac{1-z^2}{(2-z^2)^{1/2}} \, .
\end {eqnarray}
Note that $m_\infty \simeq q_{\rm max} \sin\theta$ for small
$\sin\theta$.  $f_1(z)$ runs from 0 to $1/\sqrt2$ and $f_2(z)$ from
$1/\sqrt2$ to 0 as $z$ runs from 0 to $\infty$.
For $q\ll m_\infty$, one gets $\gamma \simeq q$ in this
small $\sin\theta$ limit.  The maximum growth rate for a given
small $\sin\theta$ occurs at
$q \simeq m_\infty^2 (\frac23 \sin^2\theta)^{1/4}.$

For $\sin\theta=0$,
the result becomes simply
\begin {equation}
   \frac{\gamma}{m_\infty} = f_1\left(\frac{q}{m_\infty}\right)
   \qquad
   (\sin\theta = 0) ,
\label {eq:Bsmalltheta}
\end {equation}
which, as $q\to\infty$, approaches the maximum growth rate for the
planar problem of $\gamma = m_\infty/\sqrt2$.
By symmetry the behavior of $\Pi^{xx}$ must
be the same as that of $\Pi^{yy}$ at $\sin\theta = 0$,
and so there are two rather
than one magnetically
unstable modes at exactly $\sin\theta = 0$.


\subsubsection {Thick planar distributions and related matters}
\label {sec:etasmear}

\begin{figure}
\includegraphics[scale=0.40]{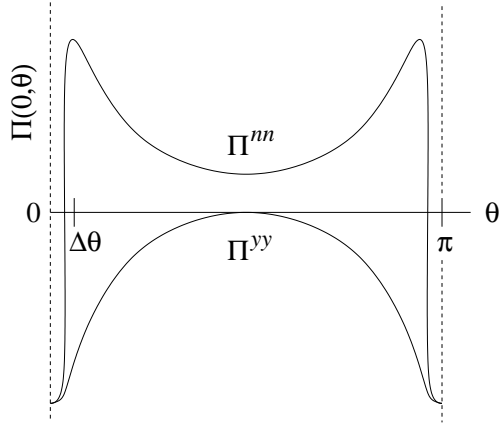}
\caption{%
    Qualitative behavior of $\Pi^{\hat\n\hat\n}(0,\hat\hat\q)$ and
    $\Pi^{yy}(0,\hat\q)$ vs.\ $\hat\q$
    for a planar distribution with small thickness $\Delta\theta$.
    The curves for $\sin\theta \gg \Delta\theta$ depict the formulas
    of (\ref{eq:Piresult}).
    \label{fig:thick}
    }
\end{figure}

One can now ask what happens if the planar distribution has a tiny
thickness, so that $\delta(p_z)$ is replaced by something with a small
but finite width $\Delta p_z$.  Let $\Delta\theta \sim (\Delta p_z)/p$ be the
angular width of the distribution, where $p$ is the momentum scale that
dominates $f(\p)$, and assume $\Delta\theta \ll 1$.
The effect this will have is to smear out the
$\theta$ dependence of previous results over $\Delta\theta$.  In
particular, $\Pi^{yy}(0,\hat\q)$ will now be finite for $\hat\q$ in
the $\pm z$ direction, with (\ref{eq:qmax}) replaced at
$\theta = 0$ by
\begin {equation}
   q_{\rm max}(\theta{=}0) \sim \frac{m_\infty}{\Delta\theta} \, .
\end {equation}
Also, there will now be two unstable modes for
$\sin\theta \lesssim \Delta\theta$ rather than just at $\sin\theta=0$.
The qualitative behavior of $\Pi^{xx}(0,\hat\q)$ and $\Pi^{yy}(0,\hat\q)$
as a function of $\theta$ is shown in Fig.\ \ref{fig:thick}.

Though the preceding is a correct analysis of ``magnetic'' instabilities
as defined by Condition 1 of Sec.\ \ref{sec:criteria},
it is qualitatively a bit misleading.
Instead of focusing on the zero frequency self-energy used in
Condition 1, let us make some very rough qualitative arguments
about the self-energy at
the non-zero frequency $\omega=i\gamma$ corresponding to the
growing unstable mode.
The origin of the formal divergence of the result (\ref{eq:piyy})
for $\Pi^{yy}(0,\q)$ as $\theta\to0$ is in the denominators
$(-\eta+\hat\p\cdot\hat\q-i\epsilon)^2$ for the
zero-thickness ($\Delta\theta = 0$) planar distribution.
For $\eta=0$, this is negative (destabilizing) only when
$\hat\p\cdot\q = 0$.  For $\eta=i\gamma/q$, it is instead destabilizing
when $\hat\p\cdot\q \lesssim \gamma/q$.  For $\q$ close to the
$z$ axis, the average value of $(\hat\p\cdot\hat\q)^2$ is of order
$\theta^2$.  The qualitative effect of non-zero $\gamma$ is therefore
similar to thickening out the planar distribution by
$\Delta\theta \sim \gamma/q$.  For small $\theta$, we have
found magnetic instabilities with $\gamma/q \sim 1$ when
$q \ll q_{\rm max}$, which is therefore similar to $\Delta\theta \sim 1$.
We therefore expect that for small $\theta \ll 1$ and $q \ll q_{\rm max}$,
the smearing
due to non-zero $\eta$ should be enough to allow an $x$-polarized unstable
solution that looks just like the $y$-polarized solution we have already
found.  Indeed, we shall discover just such a solution in the next section,
where we turn to the ``electric'' instabilities associated with Condition 2
of Sec.\ \ref{sec:criteria}.


\subsection {``Electric'' instabilities in the ultra-relativistic limit}

\subsubsection {\boldmath$\omega=0$ electric stability analysis}

Let us look for electric instabilities by checking Condition 2 of
Sec.\ \ref{sec:criteria}.  First consider $\Pi^{00}(0,\q)$.
Starting from the general expression (\ref{eq:pi}) for the self-energy
and integrating by parts, one finds
\begin {equation}
   \Pi^{00}(0,\hat\q) = e^2 \int_{\p} f(\p) \left[
        \frac1{vp} \left(-1 + \frac{1}{(\hat\p\cdot\hat\q-i\epsilon)^2}
                   \right)
        + \frac1{v^2}\,\frac{dv}{dp} \right] .
\label {eq:pi00}
\end {equation}
Let's focus on the case $\sin\theta\not=0$.
For the planar distribution function, the angular integral of
$(\hat\p\cdot\hat\q-i\epsilon)^{-2}$ over $\hat\p$ in the $xy$ plane
gives zero, and one finds
the $\hat\q$-independent result
\begin {equation}
   \Pi^{00}(0,\hat\q)
    = e^2 \int_{\p} f(\p) \left[
        - \frac1{vp}
        + \frac1{v^2}\,\frac{dv}{dp} \right] .
\end {equation}
This result gives zero in the non-relativistic limit.  We will focus on
the ultra-relativistic limit, where
\begin {equation}
   \Pi^{00}(0,\hat\q) = - m_\infty^2 \, ,
\end {equation}
which is negative and so stabilizing.

But now we need to examine $\Pi^{0i}(0,\hat\q)$.
{}From $y \to -y$ reflection symmetry, one has $\Pi^{0y}(0,\hat\q) = 0$ and 
{}from the Ward identity, $\Pi^{0\hat\q}(0,\hat\q) = 0$.  It remains
to calculate $\Pi^{0\hat\n}(0,\hat\q)$.
One can use (\ref{eq:Pi0i}), but we find it simpler to proceed similarly
to earlier calculations with the planar distribution.
Integrating (\ref{eq:pi}) for the self-energy by parts,
\begin {equation}
    \Pi^{0i}(0,\hat\q) = e^2 \int_{\p} \frac{f(\p)}{p} \,
         \frac{\hat v^i - \hat q^i \hat\v\cdot\hat\q}
              {(\hat\v\cdot\hat\q-i\epsilon)^2} \, .
\end {equation}
In particular, we can write
\begin {equation}
   \Pi^{0\hat\n}(0,\hat\q) = e^2 \int_{\p} \frac{f(\p)}{p} \,
        \left\langle
           \frac{(-\cos\theta\cos\phi)}{(\sin\theta\cos\phi-i\epsilon)^2}
        \right\rangle_\phi
   = - \frac{i \cos\theta}{\sin^2\theta} \> e^2 \int_\p \frac{f(\p)}{p} .
\end {equation}
In the ultra-relativistic limit, this is just
\begin {equation}
   \Pi^{0\hat\n}(0,\hat\q)
   = - i m_\infty^2 \, \frac{\cos\theta}{\sin^2\theta} .
\end {equation}

Now we can apply Condition 2 of Sec.\ \ref{sec:criteria}, which indicates
an electric instability if
\begin {equation}
      q^2 - \Pi^{00}(0,\hat\q)
      + \Pi^{0i}(0,\hat\q) \left[q^2+{\bm\Pi}(0,\q)\right]^{-1}_{ij}
        \Pi^{j0}(0,\hat\q)
      < 0 .
\label {condition2z}
\end {equation}
Making use of the result (\ref{eq:Pinnstatic})
for $\Pi^{\hat\n\hat\n}(0,\hat\q)$,
this condition becomes, for the ultra-relativistic planar distribution,
\begin {equation}
   q^2 + m_\infty^2
    - \frac{m_\infty^4 \cot^2\theta}{q^2\sin^2\theta + m_\infty^2}
   < 0 .
\label {eq:Econdition}
\end {equation}
The left-hand side is minimized for $q = 0$, in which case the inequality
gives $1-\cot^2\theta < 0$.  So, there will be some mode of electric
instability whenever $\hat\q$ lies within 45 degrees of the $z$ axis.%
\footnote{
   This instability corresponds to poles in $\Delta_G$ of Ref.\
   \cite{strickland} in the $\xi\to+\infty$ limit.
}
For a given $\theta$, the maximum $q$ which gives an instability is
given by (\ref{eq:Econdition}) as
\begin {equation}
  q_{\rm max} = m_\infty \, \left[
     \frac{\cos\theta \, (4+\cos^2\theta)^{1/2}
           -\sin^2\theta-1}
     {2\sin^2\theta}
  \right]^{1/2} .
\end {equation}

Since $-\Pi^{00}(0,\hat\q) > 0$, which is stabilizing,
the ``electric''
instability in this case is not due to the simple picture of electric
instabilities discussed in Sec.\ \ref{sec:physical} but instead depends on
the coupling of charge and current fluctuations through $\Pi^{i0}$.
An example of a distribution with an electric instability whose origin
does have the simple interpretation of Sec.\ \ref{sec:physical}
[having $\Pi^{i0}(0,\hat\q)=0$]
is given in Appendix \ref{app:line}.


\subsubsection {Electric instability growth rate}

\begin{figure}
\includegraphics[scale=0.40]{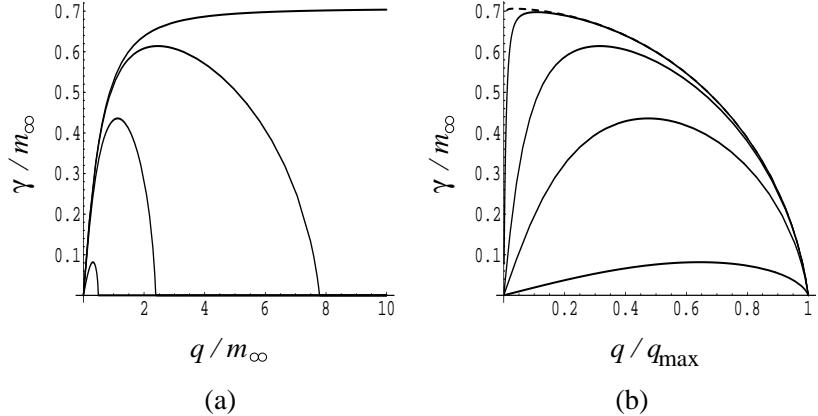}
\caption{%
    The electric instability 
    growth rate vs.\ $q$ for the planar
    distribution for various values of $\theta$.
    {}From the top down, the values are $\sin\theta$ = 0.01, 0.1, 0.3, 0.65.
    The small dashed line at the top of (b) shows the $\sin\theta\to0$ limit.
    There are no solutions for $\sin\theta > 1/\sqrt2$.
    \label{fig:resultsXZ}
    }
\end{figure}

In Appendix \ref{app:electric}, we briefly summarize the calculation of
the dispersion relation for polarizations of $\A$ in the $xz$ plane
($\hat\q\hat\n$ plane), just as (\ref{eq:dispyy})
gave the dispersion relation for the $y$ polarization.
The result is%
\footnote{
  By re-arranging terms algebraically and squaring, one can convert this
  into a quintic equation in $\eta^2$, a subset of whose roots are
  the solutions
  to (\ref{eq:dispxz}).
}
\begin {multline}
   \cos^2\theta
     \left(1-\frac{\sin^2\theta}{\eta^2}\right)^{-1}
     \left[ \left(1-\frac{\sin^2\theta}{\eta^2}\right)^{-1/2} - 1 \right]
\\
   + \eta^2 \bigl[ 1 + (1-\eta^2) Q^2 \bigr]
     \left[ \left(1-\frac{\sin^2\theta}{\eta^2}\right)^{-1/2} - 1
              - Q^2 \sin^2\theta \right]
   = 0 ,
\label{eq:dispxz}
\end {multline}
where we have introduced the dimensionless variable $Q \equiv q/m_\infty$.
The unstable solutions are plotted in Fig.\ \ref{fig:resultsXZ} for
various choices of $\theta$.
Unlike the magnetic case, there are no solutions for angles more than
$45^\circ$ from the $z$ axis.
For any given $\q$, the maximum instability
occurs for $\sin\theta\to 0$, which is a limit in which the unstable
solution to (\ref{eq:dispxz}) becomes
\begin {equation}
   \frac{\gamma}{m_\infty} \to
   f_1\left(\frac{q}{m_\infty}\right) ,
\label {eq:gammaE}
\end {equation}
which is just the same as (\ref{eq:Bsmalltheta}) for the magnetic case.
One may check (see Appendix \ref{app:electric}) that the corresponding
polarization of $\E$ is in the $x$ direction.  This is just
the solution previously predicted for small $\sin\theta$ in
Sec.\ \ref{sec:etasmear}.

One may also explore taking $\sin\theta\to0$ for $q\sim q_{\rm max}$
as in Fig.\ \ref{fig:resultsXZ}b (rather than $q$ fixed).
In this limit, both $q_{\rm max}$ and the functional form for
$\gamma$ in terms of $q/q_{\rm max}$ are different than in the
magnetic case.
We do not have a simple closed-form expression for this limit in the
electric case.

Summarizing our results for all types of instability (electric and
magnetic) for the planar distribution (\ref{eq:fplanar}), we have found
that there are two polarizations of instability for $\q$ within 45 degrees
of the $z$ axis and one polarization otherwise.  The most unstable modes
are those with $\q$ close to or along the $z$ axis.


\section {Non-Abelian example of saturating the instability}
\label{sec:twirl}

We now give a concrete example of one particular way in which non-Abelian
interactions between soft fields can halt the continuing growth of
the amplitude of an unstable mode.  Specifically, we will show that there
exist static non-linear wave solutions to the soft field equations of
motion which, in the linear approximation, would correspond to unstable modes.
The soft field amplitudes will be of order $q/g$, as in
(\ref{eq:soft_saturate}).
We will assume that the soft modes of interest
have momentum $q$ much smaller than the momentum $p$ of typical hard
particles, so that $gA/p \ll 1$.  This means that the interactions of
soft fields with each other must be treated non-perturbatively, but the
interaction of soft fields with individual hard particles is perturbative.
We can therefore ignore non-linear interactions induced by the hard
particles and focus just on the perturbative
self-energy $\Pi$ they induce for the soft fields.
The relevant soft field equations are then the non-Abelian Maxwell
equations with $j^\mu$ given as before by the perturbative self-energy:
\begin {equation}
   D_\nu F^{\mu\nu} \simeq \Pi^{\mu\nu} A_\nu ,
\label{eq:MaxwellNL}
\end {equation}
where $F^{\mu\nu}$ is the full non-Abelian field strength and
$D$ is the covariant derivative $\partial - i g A$ in the adjoint
representation.
The only assumption we will make about the hard particle distribution is
that it is axi-symmetric about the $z$ axis so that (\ref{eq:MaxwellNL})
also has this symmetry.


\subsection {A solution}

Consider any SU(2) subgroup of the gauge group.  Within this subgroup,
we take the group structure constants to be the SU(2) ones
$f^{abc} = \epsilon^{abc}$.  Working in Lorentz gauge
($\partial_\mu A^\mu = 0$), our non-Abelian Maxwell equation is then
\begin {equation}
   \Box A^{\mu a} + g \epsilon^{abc} \partial_\nu(A^{\nu b}A^{\mu c})
   + g^2 \epsilon^{bac} \epsilon^{cde} A_\nu^b A^{\nu d} A^{\mu e}
   = \Pi^{\mu\nu} A_\nu^a \, .
\label {eq:MaxwellNL2}
\end {equation}
We will look for solutions with the ansatz
\begin {equation}
  A^a_\mu(\x) =
  \begin {cases}
     {\cal A} \, R_{a\mu}(qz), & \mbox{$a=1$ or $2$, and $\mu=1$ or $2$}; \\
     0                         & \mbox{otherwise} ,
  \end {cases}
\label {eq:twirl}
\end {equation}
where $a$ is the adjoint color index and $R_{ij}(\theta)$ is the
$2\times2$ rotation matrix
\begin {equation}
   R_{ij}(\theta) =
   \begin {pmatrix}
      \phantom{-}\cos\theta & \sin\theta \\
      -\sin\theta & \cos\theta 
   \end {pmatrix}
   .
\end {equation}
The derivative in the second term of (\ref{eq:MaxwellNL2}a) then vanishes.
Using $R_{bi}R_{di} = \delta_{bd}$, one then obtains
\begin {equation}
   (-q^2 - 2 g^2 {\cal A}^2) {\cal A} \, R_{a\nu}(qz)
   = \Pi_\perp(0,q{\bm e}_z) \, {\cal A} \, R_{a\nu}(qz) \, ,
\end {equation}
where $\Pi_\perp \equiv \Pi^{xx} = \Pi^{yy}$.
We have a solution if
\begin {equation}
   {\cal A}^2 = - \frac1{2g^2} \bigl[ q^2 + \Pi_\perp(0,q{\bm e}_z) \bigr]
\end {equation}
is positive.  
That will occur exactly when $q^2 + \Pi_\perp(0,q{\bm e_z}) < 0$,
which is precisely Condition 1 of Sec.\ \ref{sec:criteria}
for a magnetic instability associated
with small-amplitude waves with the same $\q = q {\bm e}_z$.
Note that in our static non-linear solution, ${\cal A} \propto 1/g$
and so is non-perturbatively large.

The energy density stored in the soft field is
$\sim q^2 {\cal A}^2$.  Since ${\cal A} \sim q/g$ and $q \sim m_\infty$,
the energy density in the soft field is $\sim m_\infty^4/g^2$.  This is
the same energy density we would obtain if all potentially unstable
modes (of order half of all modes with $ q < m_\infty$) carried
occupation number $\sim 1/g^2$.  The latter is the situation we actually
expect, when the instability saturates.


\subsection {A simple analog in scalar theory}

To motivate the ansatz (\ref{eq:twirl}), it is instructive to note that
there are analogous static solutions of simple $\phi^4$ theory
of a complex scalar $\phi$ with a Mexican hat potential
\begin {equation}
   V(\phi) = - \mu^2 |\phi|^2 + \half \lambda |\phi|^4 .
\end {equation}
Here the equation of motion is
\begin {equation}
   \Box\phi - \lambda |\phi|^2 \phi = -\mu^2 \phi ,
\end {equation}
and
$-\mu^2$ plays the role of the destabilizing negative $\Pi$.
By taking the ansatz
\begin {equation}
   \phi(\x) = {\cal P} e^{i q z},
\label{eq:phitwirl}
\end {equation}
where $e^{i\theta}$ is like the rotation matrix $R(\theta)$, one obtains
\begin {equation}
   {\cal P}^2 = - \frac1{\lambda} [q^2 - \mu^2] ,
\end {equation}
which gives a static solution for $q < \mu$.
These $q$'s correspond to what would be the unstable modes in a
small fluctuation analysis about $\phi=0$.

The simplifications attending the use of a ``helical'' ansatz that rotates
with $\theta = qz$, like the two ansatzes above,
were also used in Ref.\ \cite{berger&davidson}
to find a special case of non-linear
solutions in the (qualitatively different) Abelian gauge theory problem.

As far as we know, the scalar solution, like the non-Abelian one preceding it,
is not relevant to any realistic physical situations.%
\footnote{
  For $q > \mu$, one can alternatively find
  {\it non}-static solutions of the scalar
  theory of the form
  $\phi(\x,t) = {\cal P} e^{i \omega t}$.
  As far as we know, these are not useful
  for anything either.
  A broader class of related solutions for the quark-gluon plasmas
  has been investigated for stable, equilibrium plasmas in
  Ref.\ \cite{iancu}.
}
In typical applications
where unstable fluctuations are growing, one will have an entire spectrum
of different growing modes which will interact with each other non-linearly
once they grow large enough.  In the scalar problem, we would not expect
a random spectrum of initial small fluctuations about $\phi=0$ to grow
into a solution like (\ref{eq:phitwirl}).  Instead, we expect this system
to equilibrate, dissipating its energy into small random fluctuations
about a minimum of the potential energy.  Similarly, though our
non-Abelian solution gives an example of how non-Abelian interactions
can saturate the growth of instabilities, we expect the actual situation
to be much more complicated.

As a final note on how things can be very different when only a limited
number of modes are involved, consider a different ansatz we could have
considered for the gauge theory problem
that only considers modes with a single (adjoint) color:
\begin {equation}
   A_\mu^a(\x) = {\cal A}(\x) \, \delta^{a1} \varepsilon_\mu .
\end {equation}
With this ansatz, the non-Abelian non-linear equations (\ref{eq:twirl}) reduce
to the Abelian linear equations (\ref{eq:Maxwell3}).
In this unnaturally idealized situation, the
growth of instabilities would not be halted by soft
interactions but, just like the Abelian case,
would continue to grow until there were non-perturbative
effects on the hard particles.  
We suspect that such a configuration in a non-Abelian setting would be
unstable to the formation of large fields in other color directions, but
we have not shown this explicitly.


\section{Conclusion}
\label{sec:conclusion}

We have seen that QCD plasma instabilities play an important role in the
idealized theoretical limit of high energies and small coupling constants and
drastically modify the bottom-up thermalization scenario.  We have also seen
reasons why the saturation of these instabilities should be qualitatively
different from the Abelian case.
An important goal for future work is to
understand what replaces the original bottom-up scenario and determine in
detail how these instabilities parametrically affect the rate of
equilibration.


\begin{acknowledgments}

We would like to thank Larry Yaffe and
Eugene Kolomeisky for useful conversations.
This work was supported, in part,
by the U.S. Department of Energy under Grant No.~DE-FG02-97ER41027, and
by a McGill University startup grant.

\end{acknowledgments}

\appendix


\section{Sufficient conditions for instability}

In order to analyze the existence of instabilities, it will be
convenient, initially, to phrase the discussion solely in terms of
$\Pi^{ij}$.  Note that $\Pi^{00}$ and $\Pi^{0i}$ can be converted using the
the Ward identity (\ref{eq:ward}).
We will follow \Mrowczynski\ and others
by rewriting the linearized Maxwell equation (\ref{eq:Maxwell3})
in the form%
\footnote{
  This equation is gauge invariant in Abelian theory.
  Gauge invariance in non-Abelian theory is not manifest simply
  because we are studying instabilities by linearizing in the fields.
  The equation is gauge invariant up to non-linear terms that have been
  neglected.
}
\begin {equation}
   D^{ij}(\omega,\q) \, E^j \equiv
   [(-\omega^2+q^2) \delta^{ij} - q^i q^j + \Pi^{ij}(\omega,\q)] E^j
   = 0 .
\label {eq:D}
\end {equation}
One quick way to obtain this is to take the time derivative of
(\ref{eq:Maxwell3}) in $A^0=0$ gauge with $\mu=i$.
The question of the existence of instabilities becomes whether
$D^{ij}(\omega,\q)$ ever has a zero eigenvalue for some $\omega$ with
$\Im\omega > 0$.

\subsection{Condition 1}
\label{app:condition1}

In searching for solutions to the dispersion relation with complex
$\omega$, it will be convenient to drop the $i\epsilon$ prescription
in the self-energy (\ref{eq:pi}) and simply absorb it into the value of
$\omega$:
\begin {equation}
   \Pi^{ij}(\omega,\q) =
   e^2 \int_\p
     \frac{\partial f(\p)}{\partial p^k}
     \left[ -v^i \delta^{kj}
            + \frac{v^i v^j q^k}{-\omega+\v\cdot\q} \right] .
\label{eq:piw}
\end {equation}
Changing integration variables $\p \to -\p$ in this expression then
implies $\Pi^{ij}(\omega) = \Pi^{ij}(-\omega)$.
Complex conjugation then yields%
\footnote{
  For $\omega = \omega_0 + i\epsilon$ with $\omega_0$ real, this
  implies
  $\Pi^{ij}_{\rm ret}(\omega_0) = - \Pi^{ij}_{\rm adv}(-\omega_0)
   = - [\Pi^{ij}_{\rm ret}(-\omega_0)]^*$,
  where ``ret'' and ``adv'' denote the self-energies for
  retarded and advanced Green's functions.
}
\begin {equation}
   \Pi^{ij}(\omega) = [\Pi^{ij}(-\omega^*)]^* .
\label {eq:conjugate}
\end {equation}
In particular, consider pure positive imaginary $\omega$, and write
$\omega=i\gamma$ with $\gamma$ real.
Then (\ref{eq:conjugate}) implies $\Pi^{ij}(i\gamma)$ is a real matrix.
Since it is also a symmetric matrix, it therefore has real eigenvalues.

\begin{figure}
\includegraphics[scale=0.60]{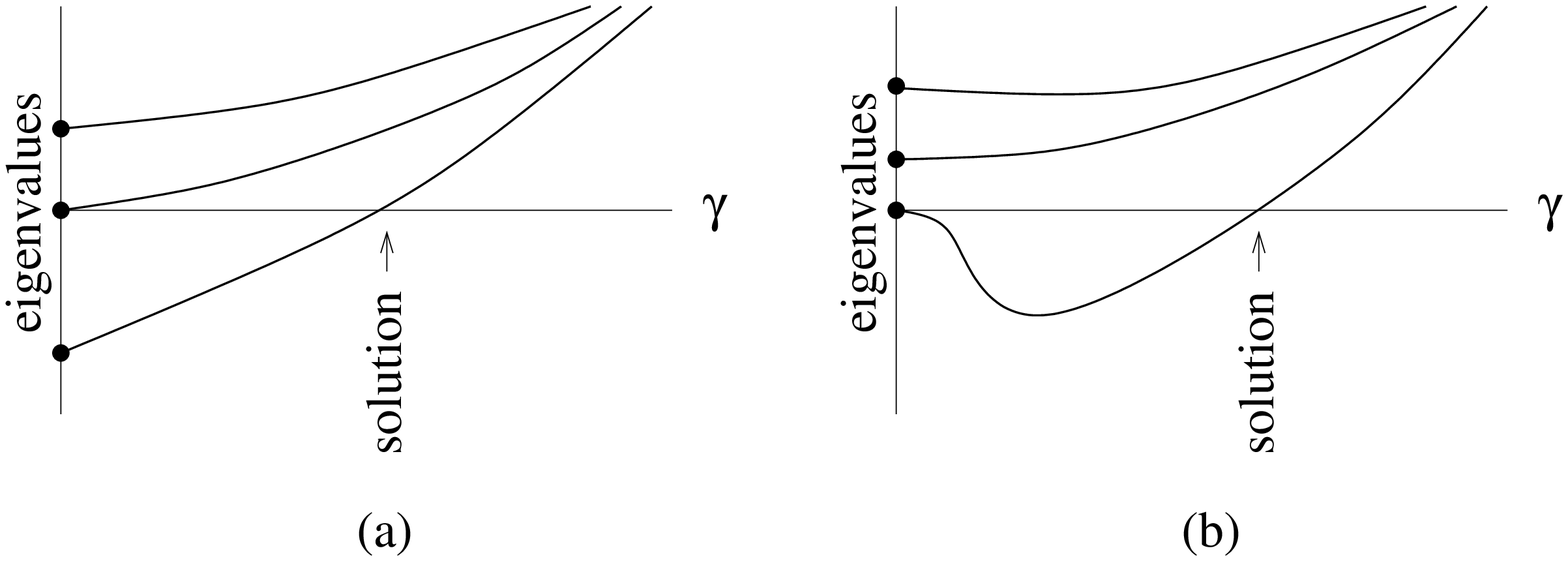}
\caption{%
    \label{fig:continuity}
    Examples of the continuity of eigenvalues of $D^{ij}(i\gamma,\q)$
    with $\gamma$, depicting (a) one magnetic instability [Condition 1],
    and (b) one electric instability [Condition 2].
    }
\end{figure}

The expression (\ref{eq:piw}) for $\Pi^{ij}(\omega,q)$ is bounded in
magnitude as $\omega \to i \infty$.  In this limit, $D^{ij}$ given
by (\ref{eq:D}) then becomes
\begin {equation}
  D^{ij}(\omega,\q) \to - \omega^2 \delta^{ij} = \gamma^2 \delta^{ij} ,
\end {equation}
with eigenvalues becoming identical and going to $+\infty$ as
$\gamma \to \infty$.  On the other hand, for $\gamma=0$,
\begin {equation}
  D^{ij}(0,\q) = q^2 \delta^{ij} - q^i q^j + \Pi^{ij}(0,\q) ,
\end {equation}
which is the $3\times3$ matrix of Condition 1' of Sec.\ \ref{sec:criteria}.
Suppose $D^{ij}(0,\q)$ has one or more negative eigenvalues,
such as shown in Fig.\ \ref{fig:continuity}a.
By continuity of the real eigenvalues of $D^{ij}(0,i\gamma)$
as $\gamma$ is varied from 0 to $\infty$,
there must then be some positive $\gamma$ for
which $D^{ij}(0,i\gamma)$ has a zero eigenvalue, which corresponds
to an unstable solution.
And there must exist one such solution for each negative eigenvalue
of $D^{ij}(0,\q)$.
This proves Condition 1', which is equivalent to Condition 1.

\subsection{Condition 2}
\label{app:condition2}

There is always a zero eigenvalue of $D^{ij}(0,\q)$ associated with
the longitudinal polarization $\hat\q$, and the above argument does not
tell us whether there is a $\gamma>0$ solution $D^{ij}(i\gamma,\q)=0$
associated with the continuous evolution of this eigenmode as $\gamma$
varies from 0 to $\infty$.  However, suppose we knew that the
corresponding eigenvalue was negative for very small positive $\gamma$.
The previous eigenvalue continuity argument would then guarantee a
corresponding instability, as depicted in Fig.\ \ref{fig:continuity}b.

Let us therefore investigate the small eigenvalue $\lambda$ of
$D^{ij}(i\gamma,\q)$ for arbitrarily small and positive $\gamma$.
Let $\top$ represent the two directions transverse to $\q$.
Making use of the Ward identity (\ref{eq:ward}), the matrix
$D^{ij}(\omega,\q)$ for small $\omega$ has the form
\begin {equation}
   D(\omega,\q)
   =
   \begin {pmatrix}
      -\omega^2 + \Pi^{\hat q \hat q}(\omega,\q)
              & \Pi^{\hat q\top}(\omega,\q) \\[8pt]
      \Pi^{\top \hat q}(\omega,\q) & q^2 + \Pi^{\top\top}(\omega,\q) \\
   \end {pmatrix}
   \simeq
   \begin {pmatrix}
      -\omega^2 + \frac{\omega^2}{q^2} \, \Pi^{00}(0,\q)
              & \frac{\omega}{q} \, \Pi^{0 \top}(0,\q) \\[8pt]
      \frac{\omega}{q} \, \Pi^{\top 0}(0,\q)
              & q^2 + \Pi^{\top\top}(0,\q) \\
   \end {pmatrix}
   ,
\end {equation}
where $\Pi(0,\q)$ above should be understood (as in the main text)
as the retarded self-energy at zero frequency, $\Pi(i\epsilon,\q)$.
In the limit of small $\omega$, the small eigenvalue of such a matrix is
\begin {equation}
   \lambda \simeq
   \frac{\omega^2}{q^2} \left[
          -q^2 + \Pi^{00} - \Pi^{0\top} (q^2+\Pi^{\top\top})^{-1} \Pi^{0\top}
   \right]_{\omega=i\epsilon} ,
\label {eq:lambda}
\end {equation}
where $[q^2 + \Pi^{\top\top}]^{-1}$ is the inverse of the $2\times2$ matrix
$q^2+\Pi^{\top\top}$.  Condition 2 of Sec.\ \ref{sec:criteria} is just
the condition that the eigenvalue (\ref{eq:lambda}) be negative for
small imaginary $\omega$.


\subsection {Relation to Nyquist analysis}

In textbooks \cite{krall,davidson},
it is more usual to derive the Penrose criteria from
a Nyquist analysis rather than to use continuity arguments as
above.  However, the usual analysis becomes more complicated
than textbook examples for
general distributions $f(\p)$ and general directions $\hat\q$,
and we find the continuity argument simpler to implement for reasons
we now explain.

\begin{figure}
\includegraphics[scale=0.40]{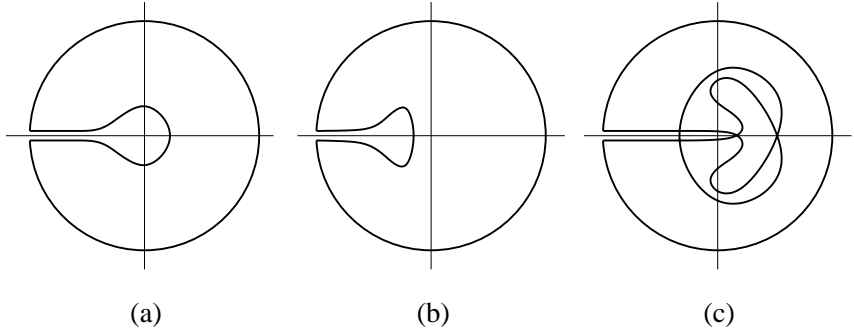}
\caption{%
    Examples of Nyquist contours $D(C)$ in the complex plane.
    The large outer circle represents
    the image of the semi-circle at infinity, and the rest is the image
    of ${\Re}+i\epsilon$.  [Diagram (c) is not just a fanciful example.
    It is also the form of the Nyquist diagram
    associated with the function $\eta^{-2}$ times the left-hand side
    of (\ref{eq:dispxz})
    for certain values of $\q$ and $\sin\theta$
    that give instability.  The single winding denotes a single associated
    unstable mode.]
    \label{fig:nyquist}
    }
\end{figure}

The textbook derivations work whenever one can use symmetry to reduce
the matrix dispersion relation (\ref{eq:D}) to single-component
dispersion relations.  If one is looking for solutions to a single
component relation of the form
\begin {equation}
   D(\omega,\q) \equiv -\omega^2+q^2+\Pi(\omega,\q) = 0 ,
\label {eq:1component}
\end {equation}
and if $\Pi$ is an analytic function of $\omega$ in the upper half plane,
then one can count the number of zeros of $D(\omega)$ for fixed $\q$
(which gives the number of unstable solutions) using the Residue Theorem
as
\begin {equation}
   N = \oint_C \frac{d\omega}{2\pi i}\,\frac1{D} \,\frac{dD}{d\omega}
     = \oint \frac{d(\ln D)}{2\pi i} ,
\end {equation}
where the contour $C$ encloses the upper half $\omega$ plane by (i)
running infinitesimally above the real axis (which we will denote
${\Re}+i\epsilon$) and then (ii) closing in a
semi-circle at infinity.  The last form above shows that $N$ is just
the winding number about the origin of the image $D(C)$ of the curve $C$.
For $|\omega|\to\infty$, the self-energy given by (\ref{eq:pi2}) is
bounded, and so $D(\omega) \to -\omega^2$, so that the semi-circle
at infinity maps to a circle at infinity that almost closes.
Fig.\ \ref{fig:nyquist} shows various examples of possible images $D(C)$.
The winding number in all these cases is even if $\Pi(i\epsilon)$ is
positive and odd otherwise, for the following reasons.
Count the winding number by counting the number of times (and sense)
that $D(C)$ crosses the negative real axis.
The curves $D(C)$ are symmetric with respect to complex conjugation because
the conjugation property (\ref{eq:conjugate}) of the self-energy implies
$\Pi(x+i\epsilon) = \Pi^*(-x+i\epsilon)$ for real $x$, so that the image
of the half-line infinitesimally above the positive real axis must be the
conjugate of that above the negative real
axis.  This conjugation property means that, except at $\omega=i\epsilon$,
any crossing of the real axis by $D(C)$
must be at a point where the real axis is
crossed {\it twice}.  As a result, all the crossings except for
$\omega=i\epsilon$ do not affect whether the winding number is odd or
even, which is solely determined by $\Pi(i\epsilon)$.
If $\Pi(i\epsilon)$ is odd, one knows that the winding number cannot
be zero, which proves the sufficiency of the Penrose criterion.

The difficulty with the above analysis in the multi-component case
is that it relies on
$D(\omega)$ being analytic in the upper half plane.  Suppose
that in the general multi-component case of (\ref{eq:D}) one wants to
develop three {\it independent}\/ instability conditions, corresponding
to the three eigenvalues of the matrix $D$.  One could imagine
algebraically solving for the eigenvalues to get three one-component
equations.  However, solving the characteristic equation to obtain
eigenvalues introduces root singularities and branch cuts.  If any of
these appeared in the upper-half plane, it would destroy the
analyticity assumption of the Nyquist analysis.  It may be possible
to construct an analysis of this sort, but we found it easier to
avoid the issue altogether.

One may forgo finding independent conditions and simply calculate the
total number of unstable solutions by doing a Nyquist analysis with
the single equation $\det D = 0$.  However, there are not useful
corresponding Penrose criteria because the sign of $\det D(i\epsilon)$
would depend on whether there were an even or odd number of unstable
modes.

For cases where the problem can be reduced to one-component equations
like (\ref{eq:1component}), one often sees the Penrose criteria used
as {\it necessary}\/ as well as sufficient conditions for instability.
This happens, for example, in situations where one can argue that
the sign of
$\Im \Pi(x+i\epsilon)$ is the same as the sign of $x$
for real $x \not=0$,
so that the {\it only}
place $D({\Re}+i\epsilon)$ can cross the real axis is for
$\omega=i\epsilon$.  Then the winding number is either exactly zero
or one, as in Figs.\ \ref{fig:nyquist}a and b, depending on the sign of
$\Pi(i\epsilon)$.  For example, consider the case
with $\hat\q$ in the $z$ direction,
$f(\p) = F(p_z) H(p_\perp)$, and polarization in the $x$ direction.
Taking the imaginary part of
(\ref{eq:pi}) for real $\omega$ gives
\begin {equation}
   \Im \Pi^{\varepsilon\varepsilon}(\omega,\q) =
   e^2 \pi \int_\p \frac{\partial f}{\partial p_z} \, (v_x)^2 \delta(-\eta +
   v_z) \, ,
\end {equation}
where $\eta \equiv \omega/|\q|$.
The sign of this result is then just the sign of $\partial F/\partial p_z$
evaluated at $v_z=\eta$.  If $F(p_z)$ is a monotonically decreasing
function of $|p_z|$, then the sign of the result will indeed be the sign of 
$\omega$, and the Penrose condition would be necessary as well as
sufficient for that mode.

\begin{figure}
\includegraphics[scale=0.40]{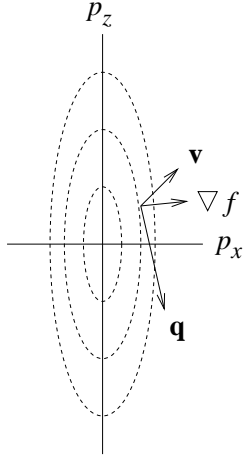}
\caption{%
    An example (shown in the $xz$ plane) of $f(\p)$, $\p$ and $\q$ for
    which $\v\cdot\grad f$ and $\v\cdot\q$ do not have the same sign.
    The ellipses indicate curves of constant $f$.
    \label{fig:negative}
    }
\end{figure}

Unfortunately, matters are more complicated when $\hat\q$ does not point
in a symmetry direction.  Suppose again that $f(\p)$ had the form
$F(p_z) H(p_\perp)$, with both $F(p_z)$ and $H(p_\perp)$ monotonically
decreasing functions of $|p_z|$ and $p_\perp$.
One could easily make a similar argument as above
if $\Im\Pi^{ij}(\omega,\q)$ were a positive definite matrix for positive
real $\omega$.
One can check positive definiteness by checking
$\pol^i \Im\Pi^{ij} \pol^j > 0$ for arbitrary real spatial
polarizations $\bpol$.
Eq.\ (\ref{fig:nyquist}) gives a result of the form
\begin {equation}
   \varepsilon_i \Im[\Pi^{ij}(\omega,\q)] \varepsilon_j
   = \int_\p \mbox{[positive]} \, \delta(-\eta+\v\cdot\hat\q)
       \, \hat\q\cdot\grad_\p f .
\end {equation}
Unfortunately, $\hat\q\cdot\grad_\p f$ is not necessarily positive
when $\v\cdot\hat\q$ is positive, as shown in Fig.\ \ref{fig:negative}.
We are unaware of a general proof that the Penrose criteria are
necessary conditions for instability in such cases.

\section{Anisotropy creates instability (Condition 1-b)}
\label {app:anisotropy}

First, we will show that non-vanishing $\Pi^{ij}(0,\hat\q)$ implies
magnetic instabilities.  Start by considering the trace of the spatial
part of (\ref{eq:pistatic}), and then spatially average that trace over
the direction of $\hat\q$:
\begin {equation}
   \bigl\langle \Pi^{ii}(0,\hat\q) \bigr\rangle_{\hat\q} =
   e^2 \int_\p
     \frac{\partial f(\p)}{\partial p^k}
     \left[ -v^k
            + v^2 \left\langle \frac{\hat q^k}{\v\cdot\hat\q-i\epsilon}
              \right\rangle_{\!\hat\q}
     \right] .
\end {equation}
It is easy to see that
\begin {equation}
   \left\langle \frac{\hat\q}{\v\cdot\hat\q-i\epsilon}
              \right\rangle_{\!\hat\q}
   = \frac{\v}{v^2}
\end {equation}
because the left-hand side (i) gives unity when dotted with $\v$, and
(ii) must be proportional to $\v$ by rotational invariance.
We therefore obtain
\begin {equation}
   \bigl\langle \Pi^{ii}(0,\hat\q) \bigr\rangle_{\hat\q} = 0 .
\end {equation}
But the trace $\Pi^{ii}$ is the sum of the eigenvalues of $\Pi^{ij}$.
What we have shown is that the eigenvalues of $\Pi^{ij}(0,\hat\q)$,
averaged over all eigenvalues and all directions of $\hat\q$, average to
zero.  Therefore, either all eigenvalues in all directions are
identically zero, or some eigenvalue in some direction must be negative
(implying an instability).
In particular it is a necessary condition for stability that
the trace $\Pi^{ii}(0,\hat\q)$ should vanish for all $\hat\q$.

Now we show that anisotropic ${\cal M}(\hat\p)$, defined by
(\ref{eq:calM}), implies non-vanishing trace $\Pi^{ii}(0,\hat\q)$ for
some $\hat\q$ and therefore implies the existence of instabilities.
Taking the trace and setting $\omega=0$ in (\ref{eq:pi2}), we have
\begin {equation}
   \Pi^{ii}(0,\q)
   = e^2 \int_\p f(\p)
       \frac{v}{p} \left[
          1
          + \frac{1}{(\hat\p\cdot\hat\q-i\epsilon)^2}
         \right] .
\end {equation}
Note that the factor in brackets does not depend on the magnitudes $p$ and
$v$ of $\p$ and $\v$ but only on their common direction $\hat\p$.
Performing the integral over $p$ gives a factor of ${\cal M}(\hat\p)$
defined by (\ref{eq:calM}):
\begin {equation}
   \Pi^{ii}(0,\hat\q) =
   \left\langle {\cal M}(\hat\p) \,
     \left[ 1 + \frac{1}{(\hat\p\cdot\hat\q-i\epsilon)^2} \right]
   \right\rangle_{\hat\p} .
\label {eq:PiY}
\end {equation}
Now decompose ${\cal M}$ into spherical harmonics:
\begin {equation}
   {\cal M}(\hat \p) = \sum_{lm} a_{lm} \, Y_{lm}(\hat\p) .
\end {equation}
By rotational invariance, Eq.\ (\ref{eq:PiY}) will generate an
angular dependence proportional to $Y_{lm}(\hat\q)$ for each term
proportional to $Y_{lm}(\hat\p)$.  So
\begin {equation}
   \Pi^{ii}(0,\hat\q) = \sum_{lm} \kappa_l \, a_{lm} \, Y_{lm}(\hat\q) ,
\end {equation}
where
\begin {equation}
   \kappa_{l} \equiv 4\pi \left\langle
     Y_{lm}^*(\hat\q)
     \left[ 1 + \frac{1}{(\hat\p\cdot\hat\q-i\epsilon)^2} \right]
     Y_{lm}(\hat\p)
   \right\rangle_{\hat\p,\hat\q}
\label{eq:kappa}
\end {equation}
does not depend on $m$, again because of rotational invariance.

The only way that (\ref{eq:PiY}) can vanish for all $\hat\q$ is if
all of the $a_{lm} \kappa_l$ vanish.
If ${\cal M}(\hat\p)$ is anisotropic, then one of the $a_{lm}$ must
be non-zero for $l>0$, and it must be an even $l$ because of our
universal assumption that $f(\p)$ is parity symmetric.
If we knew that $\kappa_l \not=0$
for all even $l > 0$, we would then know that $\Pi^{ii}(0,\hat\q)$
cannot vanish for anisotropic ${\cal M}(\hat\p)$.
Therefore we turn to the evaluation of $\kappa_l$.

To evaluate $\kappa_l$, first choose $m=0$ in (\ref{eq:kappa}).
Next, use Wigner $D$ functions to re-express the spherical
harmonic $Y_{l0}(\hat\p)$ defined with respect to a fixed $z$ axis
in terms of spherical harmonics $Y^{(\hat\q)}_{lm'}(\hat\p)$
defined with respect to the direction $\hat\q$:
\begin {equation}
   Y_{lm}(\hat\p) = \sum_{m'} D_{mm'}^l(\hat\q)\,Y_{lm'}^{(\hat\q)}(\hat\p).
\end {equation}
Let $\theta$ be the angle between $\hat\p$ and $\hat\q$ and note that
\begin {equation}
   \left\langle \left[1 + \frac{1}{(\cos\theta-i\epsilon)^2}\right]
   Y_{lm'}^{(\hat\q)}(\hat\p)
   \right\rangle_{\hat\p}
\end {equation}
vanishes unless $m'=0$.
So
\begin {multline}
   \left\langle
     \left[ 1 + \frac{1}{(\hat\p\cdot\hat\q-i\epsilon)^2} \right]
     Y_{l0}(\hat\p)
   \right\rangle_{\hat\p}
   =
   D^l_{00}(\hat\q)
   \left\langle
     \left[ 1 + \frac{1}{(\hat\p\cdot\hat\q-i\epsilon)^2} \right]
     Y^{(\hat\q)}_{l0}(\hat\p)
   \right\rangle_{\hat\p}
\\
   =
   Y_{l0}(\hat\q)
   \left\langle
     \left[ 1 + \frac{1}{(\cos\theta-i\epsilon)^2} \right]
     P_l(\cos\theta)
   \right\rangle_{\hat\p} ,
\end {multline}
which gives
\begin {equation}
   \kappa_l =
   \frac12
   \int_{-1}^1 dx \> \left[1 + \frac{1}{(x-i\epsilon)^2}\right] P_l(x) ,
\end {equation}
where $P_l(x)$ is the $l$-th Legendre polynomial.
This integral gives%
\footnote{
  This may be derived by an appropriate analytic continuation of
  Eq.\ (7.126.1) of Ref.\ \cite{gr}.
}
\begin {equation}
   \kappa_l =
   \delta_{l0} -
   \frac{(-)^{\frac{l}2} \sqrt\pi \left(\frac{l}2\right)!}
        {\Gamma(\frac{l}2+\frac12)}
   = 
   \delta_{l0} -
   \frac{(-)^{\frac{l}2} \, l !!}{(l-1)!!}
\end {equation}
for even $l$, which indeed does not vanish for any even $l > 0$.

This completes the proof that anisotropic ${\cal M}(\hat\p)$ implies
magnetic instabilities.  What about the converse?
Suppose ${\cal M}(\hat\p)$ is isotropic.
In the ultra-relativistic limit ($v=1$), we can perform the
radial $p$ integral of the expression (\ref{eq:pi2}) for the
self-energy to write
\begin {equation}
   \Pi^{ij}(\omega,\q) =
   \left\langle {\cal M}(\hat\p)
          \left[
          \delta^{ij}
          - \frac{\hat q^i \hat p^j+\hat q^j \hat p^i}
                 {-\eta+\hat\p\cdot\hat\q-i\epsilon}
          + \frac{(-\eta^2+1) \hat p^i \hat p^j}
                 {(-\eta+\hat\p\cdot\hat\q-i\epsilon)^2}
         \right]
     \right\rangle_{\!\hat\p}
\end {equation}
with $\eta\equiv\omega/q$.
If ${\cal M}(\hat\p)$ is isotropic, then we can factor it out to give
\begin {equation}
   \Pi^{ij}(\omega,\q)
   = m_\infty^2 \left\langle
          \delta^{ij}
          - \frac{\hat q^i \hat p^j+\hat q^j \hat p^i}
                 {-\eta+\hat\p\cdot\hat\q-i\epsilon}
          + \frac{(-\eta^2+1) \hat p^i \hat p^j}
                 {(-\eta+\hat\p\cdot\hat\q-i\epsilon)^2}
     \right\rangle_{\!\hat\p} .
\end {equation}
All of the details
of the distribution $f(\p)$ have factored out into the single normalization
constant $m_\infty^2$.
Except for this normalization, there is no difference between the general
case of isotropic ${\cal M}(\hat\p)$ and the specific case of equilibrium
distributions.  In equilibrium, there is no instability, and so there
cannot be one for any isotropic ${\cal M}(\hat \p)$.  (One may
check this explicitly using the standard results for the equilibrium
self-energy \cite{klimov,weldon}.)


\section {Instability of the ultra-relativistic line momentum distribution}
\label{app:line}

In the bulk of this paper, we have focused on instabilities associated
with the planar (extreme oblate) distribution of (\ref{eq:fplanar}).
It is also interesting to study the other extreme of the linear
(extreme prolate) distribution of (\ref{eq:fline}):
\begin {equation}
   f(\p,\x) = F(p_z) \, \delta^{(2)}(\p_\perp)  .
\label {eq:fline2}
\end {equation}
Our standard assumption of parity symmetry is here that $F(-p_z) = F(p_z)$.
We will specialize to the ultra-relativistic case,
where we will find that the
distribution (\ref{eq:fline2}) is
always associated with electric instabilities (in
the sense of Condition 2 of Sec.\ \ref{sec:criteria}).

Consider
directions of $\hat\q$ that do not lie in the $xy$ plane,
{\it i.e.} $\theta \not= \pi/2$.
{}From (\ref{eq:pi2}), we get the self-energy
\begin {equation}
   \Pi^{ij}(\omega,\q) = e^2 \int_\p \frac{f(\p)}{p}
      \left[ \delta^{ij}
         - \frac{ \hat q^i \hat p^j +  \hat q^j \hat p^i}
                {-\eta+\hat\p\cdot\hat\q - i\epsilon}
         + \frac{(-\eta^2+1) \hat p^i \hat p^j}
                {(-\eta+\hat\p\cdot\hat\q - i\epsilon)^2}
      \right] .
\end {equation}
Naively, one may use the $\delta$ function in the distribution
(\ref{eq:fline2}) to do all but the $p_z$ integral.  Summing over
the cases $p_z < 0$ and $p_z > 0$ (corresponding to $\hat p_z = \pm 1$
for this distribution),
\begin {equation}
   \Pi^{ij}(\omega,\q) = \half \, m_\infty^2 \sum_{\pm}
      \left[ \delta^{ij}
         \mp \frac{\hat q^i \delta^{jz} + \hat q^j \delta^{iz}}
                {-\eta\pm\cos\theta - i\epsilon}
         + \frac{(-\eta^2+1) \delta^{iz} \delta^{jz}}
                {(-\eta\pm\cos\theta - i\epsilon)^2}
      \right] ,
\label {eq:piLine}
\end {equation}
with $m_\infty$ naively given by (\ref{eq:minfty}) as
\begin {equation}
   m_\infty^2 = e^2 \int_\p \frac{f(\p)}{p}
              = \frac{2e^2}{(2\pi)^3} \int_0^\infty dp_z \>
                   \frac{F(p_z)}{p_z} .
\label {eq:minftyLine}
\end {equation}

The preceding is
slightly naive if $F(0) \not=0$, since then the integral
(\ref{eq:minftyLine})
has a logarithmic, small $p_z$ divergence.
In physical situations, the
$\delta^{(2)}(\p_\perp)$ in (\ref{eq:fline2}) would have a small width
$\Delta p_\perp$, and the small $p_z$ divergence would cut off when
$p_z \sim \Delta p_\perp$.
Then $m_\infty^2$ will be finite, growing proportional to
$F(0)\ln[(\Delta\theta)^{-1}]$ as $\Delta p_\perp \to 0$, where
$\Delta\theta \sim p_\perp/\bar p$ and $\bar p$ is a characteristic
momentum scale of $F(p_z)$.
If we are only interested in results to leading order in this
logarithm (which come from $p_z \gg \Delta p$), then the equation
(\ref{eq:piLine}) is fine.%
\footnote{
   The non-relativistic case is very different.  Doing a similar computation
   of $\Pi^{00}(0,\hat\q)$ in the non-relativistic case, one finds a
   linear rather than logarithmic small $p_z$ divergence when
   $F(0) \not= 0$.  As a result, the $\delta$-function approximation
   in the distribution (\ref{eq:fline2}) is never valid for calculating
   $\Pi^{00}(0,\hat\q)$ in the non-relativistic case when $F(0) \not= 0$.
}
It should then be understood that this
and future equations in this section
are only accurate up to relative corrections
of order $1/\ln[(\Delta\theta)^{-1}]$.

As usual, let $\q$ lie in the $xz$ plane.
The $y$ direction is then an eigen-direction of $\Pi^{ij}$,
and (\ref{eq:piLine}) gives
\begin {equation}
   \Pi^{yy}(\omega,\q) = m_\infty^2 .
\label {eq:PiyyLine}
\end {equation}
There is no instability associated with this polarization.
The remaining directions give
\begin {subequations}
\label {eq:piLineResults}
\begin {eqnarray}
   \Pi^{xx}(\omega,\q) &=& m_\infty^2 ,
\\
   \Pi^{xz}(\omega,\q) &=& m_\infty^2
      \frac{\sin\theta \cos\theta}{\eta^2-\cos^2\theta} \,,
\\
   \Pi^{zz}(\omega,\q) &=& m_\infty^2
      \frac{\sin^2\theta (\eta^2+\cos^2\theta)}{(\eta^2-\cos^2\theta)^2} \,,
\end {eqnarray}
or equivalently
\begin {eqnarray}
   \Pi^{\hat\q \hat\q}(\omega,\q) &=& \eta^2 \Pi^{00}(\omega,\q)
     = m_\infty^2 \eta^2 \sin^2\theta \, 
       \frac{\eta^2+\cos^2\theta}{(\eta^2-\cos^2\theta)^2}
       \,,
\\
   \Pi^{\hat\q\hat\n}(\omega,\q) &=& \eta \Pi^{0\hat\n}(\omega,\q)
     = m_\infty^2 \eta^2 \sin\theta \cos\theta \,
       \frac{(-2+\eta^2+\cos^2\theta)}{(\eta^2-\cos^2\theta)^2}
       \,,
\\
   \Pi^{\hat\n\hat\n}(\omega,\q) &=&
     m_\infty^2
       \frac{(-1 + \eta^2)^2 \cos^2\theta + \eta^2 \sin^4\theta}
            {(\eta^2-\cos^2\theta)^2}
       \,.
\end {eqnarray}
\end {subequations}
In all cases, we have absorbed the $i\epsilon$ prescription into $\eta$ so
that $\eta$ now stands for $\eta+i\epsilon$.

We shall look for all solutions to the dispersion relations in a moment.
It is worth noting first that
$\Pi^{\hat\n\hat\n}(0,\hat\q) = m_\infty^2/\cos^2\theta$, so that
Condition 1 of Sec.\ \ref{sec:criteria}
for a magnetic instability is not satisfied for
$\theta \not= \pi/2$.
This makes qualitative sense because, for
$\hat\q$ not in the $xy$ plane, the distribution (\ref{eq:fline2})
does not give any trapped particles.
[The singularity of $\Pi^{\hat\n\hat\n}(0,\hat\q)$ for $\theta=\pi/2$
is a reflection of the
zero width of the transverse $\delta$ function in the distribution
(\ref{eq:fline2}) and will be cut off by a small width $\Delta p_\perp$.
{}From Condition 1-b, we know there must be a magnetic instability
associated with $\theta = \pi/2$.]

In contrast, $\Pi^{00}(0,\hat\q) = m_\infty^2\tan^2\theta$
and $\Pi^{0\hat\n}(0,\hat\q) = 0$ do indicate the existence of an
electric instability by Condition 2 when
\begin {equation}
   q < q_{\rm max} = m_\infty \tan\theta .
\label {eq:qmaxLine}
\end {equation}
Note that there is no instability
for $\sin\theta=0$, which is when $\q$ points in the same direction
as the particle motion.
As discussed in Sec.\ \ref{sec:physical},
the lack of an electric instability in this case is
a reflection of the fact that the speed of
ultra-relativistic particles cannot change from $v=1$.

The nice thing about the distribution (\ref{eq:fline2}) is that we
can in fact completely solve the dispersion relation in simple closed
form.  The $yy$ dispersion relation
$-\omega^2 + q^2 + \Pi^{yy}(\omega,\q) = 0$, with (\ref{eq:PiyyLine}), just
gives the two propagating solutions
\begin {equation}
   \omega^2 = q^2 + m_\infty^2 .
\label {eq:omegayyLine}
\end {equation}
For polarizations in the $xz$ plane, we turn to the $xz$ subspace of the
form (\ref{eq:D}) of the dispersion relation.  There will be a solution
when $\det D$ (taken over this subspace) vanishes.  Using the results
(\ref{eq:piLineResults}) for $\Pi$, this condition
$D^{xx} D^{zz} - (D^{xz})^2 = 0$
[or equivalently
$D^{\hat\q\hat\q} D^{\hat\n\hat\n} - (D^{\hat\q\hat\n})^2$ = 0]
can be simplified to
\begin {equation}
   Q^2 (\eta^2-\cos^2\theta)^2 (\eta^2 Q^2 - Q^2 - 1) + \sin^2\theta = 0 ,
\end {equation}
where we use the dimensionless variable
$Q \equiv q/m_\infty$.
This condition can be factorized into
\begin {equation}
   (\eta^2 - c^2 - Q^{-2})
   \bigl[ Q^2 \eta^4 - Q^2 \eta^2 (1+c^2) + (c^2 Q^2 - s^2) \bigr]
   = 0
\end {equation}
where $c \equiv \cos\theta$ and $s\equiv \sin\theta$.
There are three corresponding solutions for $\omega^2$, in addition
to (\ref{eq:omegayyLine}), which are
\begin {equation}
   \omega^2 = q^2 \cos^2\theta + m_\infty^2 ,
\end {equation}
and
\begin {equation}
   \omega^2 = \half
          \left[ q^2(1 + \cos^2\theta) \pm
            q \sin\theta \sqrt{q^2 \sin^2\theta + 4m_\infty^2} \right] .
\end {equation}
The lower sign in the last solution gives an instability ($\omega^2 < 0$)
when $q < q_{\rm max}$, as given by (\ref{eq:qmaxLine}).
This is the only instability of this distribution for $\theta \not= \pi/2$.


\section {Dispersion relation for planar distribution}
\label{app:electric}

In this appendix, we briefly describe the derivation of the dispersion
relation (\ref{eq:dispxz}) used to study electric instabilities for
the planar distribution (\ref{eq:fplanar}).
We look for solutions to the matrix dispersion relation
(\ref{eq:D}) for polarizations in the $xz$ plane by requiring the
determinant of $D$ in that subspace to vanish:
\begin {equation}
  0 = D^{xx} D^{zz} - (D^{xz})^2
    = (-\omega^2+q_z^2+\Pi^{xx})(-\omega^2+q_x^2+\Pi^{zz})
      - (-q_x q_z + \Pi^{xz})^2 ,
\label {eq:det}
\end {equation}
or equivalently
\begin {equation}
  0 = D^{\hat\q\hat\q} D^{\hat\n\hat\n} - (D^{\hat\q\hat\n})^2
    = (-\omega^2+\Pi^{\hat\q\hat\q})(-\omega^2+q^2+\Pi^{\hat\n\hat\n})
      - (\Pi^{\hat\q\hat\n})^2 .
\end {equation}
Starting from (\ref{eq:piplanarA})
and performing the integral over the angle $\phi$
of $\hat\p$ in the $xy$ plane, one obtains
\begin {eqnarray}
   \Pi^{zz} &=& m^2_\infty \, ,
\\
   \Pi^{xz} &=& \frac{c}{s} \left[
     \left(1-\frac{s^2}{\eta^2}\right)^{-1/2} - 1
   \right] m^2_\infty \, ,
\\
   \Pi^{xx} &=& \frac1{s^2} \left[
     c^2 
       \left(1-\frac{s^2}{\eta^2}\right)^{-3/2}
     - (2 c^2 - \eta^2) 
       \left(1-\frac{s^2}{\eta^2}\right)^{-1/2}
     + c^2 - \eta^2
   \right] m^2_\infty \, ,
\end {eqnarray}
where $s \equiv\sin\theta$ and $c\equiv\cos\theta$.
Combining these expressions 
gives a result for (\ref{eq:det}) that is $s^{-2}$
times the right-hand side of (\ref{eq:dispxz}).

In the limit $\sin\theta \to 0$, the matrix $D$ becomes diagonal with
\begin {eqnarray}
   D^{zz} &\to& m_\infty^2-\eta^2 q^2 ,
\\
   D^{xx} &\to&
   q^2(1-\eta^2)+\frac{m_\infty^2}{2}\left(\frac{1}{\eta^2}+1\right)
   \, .
\end {eqnarray}
In this limit, the unstable mode, given by (\ref{eq:gammaE}), corresponds to
$\E$ polarized in the $x$ direction.  Note, in contrast, that the
Penrose condition that predicted the existence of this unstable mode via
Condition 2 was associated with the $\hat\q$ eigen-direction of
$\Pi^{ij}(i\epsilon,\q)$ (see Appendix \ref{app:condition2}),
which for $\sin\theta\to 0$ is the $z$ direction.
This just reflects
the fact that the eigen-directions can change as one varies $\omega$
from $i\epsilon$ to the location $i\gamma$ of the actual unstable solution.


\newpage

\begin{center}
{\bf
        Erratum:
        QCD plasma instabilities and bottom-up thermalization \break
        [JHEP 08 (2003) 002]
}
\medskip

Peter Arnold, Jonathan Lenaghan, and Guy D. Moore
\end{center}
\bigskip

In the last sentence of the paragraph containing (3.11),
$q \simeq m_\infty^2 (\frac{2}{3}\sin^2\theta)^{1/4}$ should instead
read $q \simeq m_\infty [2/(3\sin^2\theta)]^{1/4}$.

Before eq.\ (4.1), there is a misleadingly general statement about when
one can ignore non-linear interactions induced by hard particles.
They can be ignored for gauge fields which only depend on a single
coordinate, such as $A = A(z)$ as considered subsequently, but not
in general.  See, for example, Ref.\ [36], as well as Ref. [40] below.

Eq.\ (4.1) should have an overall minus sign on the right-hand side.
In eq.\ (4.2),  the $\epsilon^{bac}$ in the $g^2$ term should be
$\epsilon^{abc}$.
Furthermore, the left-hand side should of eq.\ (4.2) should
have an additional term
$+g \epsilon^{abc} A^b_\nu (\partial^\nu A^{c\mu} - \partial^\mu A^{c\nu})$.
The last error is not merely typographic but was due to simultaneous
cosmic ray events in the brains of the three authors.
As a result,
the solution presented in section 4.1 is not in fact a solution
to the equations of motion.  This does not affect the rest of the paper.
Actual static (but unstable) solutions analogous to the non-linear
solutions of Ref.\ [36] may be found in Ref.\ [40].

The last paragraph in section 4.1 is misleadingly general.  It
applies only to cases where the typical momenta of unstable modes are
of order $m_\infty$.  For the approximately planar hard-particle
distributions
considered in most of the paper, the typical unstable momentum
scale is instead
of order $q_{\rm max} \gg m_\infty$, where $q_{\rm max}$ is given
by eq.\ (3.13).

\begin {itemize}
\item[{[40]}]
P. Arnold and J. Lenaghan,
``The abelianization of QCD plasma instabilities,''
hep-ph/0408052.
\end {itemize}

\end {document}